\documentclass[sigconf]{acmart} \graphicspath{{./images/}}

\setcopyright{none} 

\usepackage{xcolor} \usepackage{balance} \usepackage{caption}
\usepackage[normalem]{ulem} \usepackage{listings} \usepackage{subcaption}
\usepackage{graphicx} \usepackage{color}

\usepackage{algorithm} 
\usepackage{algcompatible}
\usepackage{algpseudocode} \usepackage{enumitem} \usepackage{xspace}
\usepackage{hyperref} \usepackage{booktabs}

\newcommand*{\reg}[1]{\texttt{\%#1}}
\newcommand*{\pg}[1]{\texttt{#1}-page}
\newcommand*{\pgs}[1]{\texttt{#1}-pages}

\newcommand*{\pu}[0]{\texttt{PrivUser}\xspace}

\newcommand*{\gate}[0]{\texttt{Gate}\xspace}

\newcommand*{\liblotr}[0]{\texttt{liblotr}\xspace}
\newcommand*{\lotrkmod}[0]{\texttt{lotr-kmod}\xspace}
\newcommand*{\lotrbuild}[0]{\texttt{lotr-build}\xspace}
\newcommand*{\lotrmusl}[0]{\texttt{lotr-libc}\xspace}
\newcommand*{\lotrlibc}[0]{\texttt{lotr-libc}\xspace}

\newcommand*{\privcall}[0]{\texttt{privcall}\xspace}
\newcommand*{\privcalls}[0]{\texttt{privcalls}\xspace}

\newcommand*{\ring}[1]{$R_{#1}$}
\newcommand*{\nrtor}[2]{$R_{#1}$$\nrightarrow$$R_{#2}$}
\newcommand*{\rtor}[2]{$R_{#1}$$\rightarrow$$R_{#2}$}
\newcommand*{\cg}[2]{$CG:$\ring{#1}$\rightarrow$\ring{#2}}
\newcommand*{\ncg}[2]{$CG:$\ring{#1}$\nrightarrow$\ring{#2}}

\definecolor{allowed}{RGB}{0,220,0}
\definecolor{comment}{RGB}{85,107,47}

\lstnewenvironment{code}[1]{\lstset{ language=#1,
    directivestyle={\color{black}}, emph={},
    keywordstyle=\color{blue}, basicstyle=\footnotesize\ttfamily,
    stringstyle={\color{black}}, frame=None, showstringspaces=false,
    xleftmargin=0.3cm, framexleftmargin=0.3cm,
    emphstyle={\color{blue}},
    keywordstyle=\color{black}\bfseries, 
    commentstyle={\color{blue}},
    morekeywords={LOTR_SECRET,PRIVUSER_SECRET,__privcall2,lcall,lret,asm,privcall,
      PRIV_SECRET,syscall,PRIVCALL_DEFINE,SYSCALL_DEFINE},
    escapechar={\^}, }}{}

\lstdefinelanguage {GAS} 
[x86masm]{Assembler} 
{morekeywords={subq,pushq,popq,pushl,popl,movq,movl, lret,
    rax,rdx,rcx,rbx,rsi,rdi,rsp,rbp, %
    r8,r8d,r8w,r8b,r9,r9d,r9w,r9b, %
    r10,r10d,r10w,r10b,r11,r11d,r11w,r11b, %
    r12,r12d,r12w,r12b,r13,r13d,r13w,r13b, %
    r14,r14d,r14w,r14b,r15,r15d,r15w,r15b}, morecomment=[l]{\#},
}

\lstnewenvironment{asm}[0]{\lstset{ language=GAS,
    directivestyle={\color{black}}, emph={},
    keywordstyle=\color{black}, basicstyle=\footnotesize\ttfamily,
    stringstyle={\color{blue}}, numberstyle=\tiny, frame=single,
    xleftmargin=0.3cm, framexleftmargin=0.3cm,
    emphstyle={\color{black}},
    keywordstyle=\color{black}\bfseries, 
    rulesepcolor=\color{gray}, morekeywords={LOTR\_SECRET,privcall,
      PRIV_SECRET,syscall,PRIVCALL_DEFINE,SYSCALL_DEFINE},
    commentstyle={\ttfamily\color{blue}}, }}{}

\begin{document}
\title{Lord of the x86 Rings: A Portable User Mode Privilege
  Separation Architecture on x86}

\author{Hojoon\ Lee}
\affiliation{%
  \institution{GSIS, School of Computing, KAIST}
}
\email{hojoon.lee@kaist.ac.kr}

\author{Chihyun\ Song}
\affiliation{%
  \institution{GSIS, School of Computing, KAIST}
}
\email{ian0371@kaist.ac.kr}

\author{Brent Byunghoon Kang}
\affiliation{%
  \institution{GSIS, School of Computing, KAIST}
}
\email{brentkang@kaist.ac.kr}

\vspace{8mm}

\begin{abstract}
  Modern applications are increasingly advanced and complex, and
  inevitably contain exploitable software bugs despite the ongoing
  efforts. The applications today often involve processing of
  sensitive information. However, the lack of privilege separation
  within the user space leaves sensitive application secret such as
  cryptographic keys just as unprotected as a "hello world"
  string. Cutting-edge hardware-supported security features are being
  introduced. However, the features are often vendor-specific or lack
  compatibility with older generations of the processors. The
  situation leaves developers with no portable solution to incorporate
  protection for the sensitive application component.
  
  We propose LOTRx86, a fundamental and portable approach for user
  space privilege separation.  Our approach creates a more privileged
  user execution layer called \emph{PrivUser} through harnessing the
  underused intermediate privilege levels on the x86 architecture. The
  PrivUser memory space, a set of pages within process address space
  that are inaccessible to user mode, is a safe place for application
  secrets and routines that access them. We implement the LOTRx86 ABI
  that exports the \privcall interface to users to invoke secret
  handling routines in PrivUser. This way, sensitive application
  operations that involve the secrets are performed in a strictly
  controlled manner. The memory access control in our architecture is
  \emph{privilege-based}, accessing the protected application secret
  only requires a change in the privilege, eliminating the need for
  costly remote procedure calls or change in address space.
  
  We evaluated our platform by developing a proof-of-concept
  LOTRx86-enabled web server that employs our architecture to securely
  access its private key during SSL connection and thereby mitigating
  the HeartBleed vulnerability by design. We conducted a set of
  experiments including a performance measurement on the PoC on
  \emph{both} Intel and AMD PCs, and confirmed that LOTRx86 incurs
  only a limited performance overhead.
\end{abstract}




\maketitle




\section{Introduction}
\label{sec:introduction}

User applications today are prone to software attacks, and yet are
often monolithically structured or lack privilege separation. As a
result, adversaries who have successfully exploited a software
vulnerability in an application can access to sensitive in-process
code or data that are irrelevant to the exploited module or part of
the application. Today's applications often contain secrets that are
too critical to reside in the memory along with the rest of the
application contents, as we have witnessed in the incident of
HeartBleed~\cite{preventing-heartbleed,the-matter-of-heartbleed}.


The conventional software privilege model that coarsely divides the
system privilege into only two levels (user-level and kernel-level)
has failed to provide a fundamental solution that can support
privilege separation in user applications. As a result, critical
application secrets such as cryptographic information are essentially
treated no differently than a "hello world" string in user memory
space. When the control flow of a running user context is compromised,
there is no access control left to prevent the hijacked context to
access arbitrary memory addresses.


Many approaches have been introduced to mitigate the challenging issue
within the boundaries of the existing application memory protection
mechanisms provided by the operating system. A number of work proposed
using the process abstraction as a unit of protection by separating a
program into multiple processes~\cite{ppe,privman,privtrans}. The key
idea is to utilize the process separation mechanism provided by the
OS; these work achieve privilege separation by splitting a single
program into multiple processes. However, this process-level
separation incurs a significant overhead due to the cost of the
inter-process communication (IPC) between the processes or address
space switching that incur TLB flushes. Also, the coarse unit of
separation still leaves a large attack surface for attackers. The
direction has advanced through a plethora of works on the topic. One
prominent aspect of the advancements is the granularity of
protection. Thread-level protection schemes ~\cite{wedge,smv,arbiter}
have reduced the protection granularity compared to the process-level
separation schemes while still suffering from performance overhead
from page table modifications. Shreds presented fine-grained
in-process memory protection using a memory partitioning feature that
has long been present in ARM called \emph{Memory Domains}
\cite{shreds}. However, the feature has been deprecated in the 64-bit
execution mode of the ARM architecture (\texttt{AARCH64}).

In the more recent years, a number of processor architecture revisions
and academic works have taken a more fundamental approach to provide
in-process protection; Intel has introduced \emph{Software Guard
  Extensions (SGX)} to its new x86 processors to protect sensitive
application and code and data from the rest of the application as well
as the possibly malicious kernel~\cite{intel-sgx-web,haven}. Intel
also offers hardware-assisted in-process memory safety and protection
features~\cite{intel-mpx-web,lwn-mpk} and AMD has announced the plans
to embed a similar feature to its future generations of x86
processors~\cite{amd-sme}. However, the support for the new processor
features are fragmented; not only that the features are not
inter-operable across processors from different vendors (Intel, AMD),
they are also only available on the newer processors.
Hypervisor-based application memory
protection~\cite{secage,overshadow} may serve to be a more portable
solution, considering the widespread adaption of hypervisors
nowadays. However, it is not reasonable for a developer to assume that
her users are using a virtual machine.

The situation presents complications for developers who need to
consider the \emph{portability} as well as the security of the
sensitive data her program processes. Therefore, we argue that there
is a need for an approach that provides a basis for an
\emph{in-process} privilege separation based on only the portable
features of the processor. An \emph{in-process} memory separation
should not require a complete address space switching to access the
protected memory or costly page table modifications.

In this paper, we propose a novel x86 user-mode privilege separation
architecture called \emph{The Lord of the x86 Rings} (LOTRx86)
architecture. Our architecture proposes a drastically different, yet
portable approach for user privilege separation on x86. While the
existing approaches sought to retrofit the memory protection
mechanisms within the boundaries of the OS kernel's support, we
propose the creation of a more privileged user layer called \pu that
protects sensitive application code and data from the \emph{normal}
user mode. For this objective, LOTRx86 harnesses the underused x86
intermediate Rings (Ring1 and Ring2) with our unique design that
satisfies security requirements that define a distinct privilege
layer. The \pu memory space is a subset of a process memory space that
is accessible to when the process context is in \pu mode but
inaccessible when in user mode. In our architecture, user memory
access control is \emph{privileged-based}. Therefore, the architecture
does not require costly run-time page table manipulations nor address
space switching.

We also implement the LOTRx86 ABI that exports the \privcall interface
that supports \pu layer invocation from user layer. To draw an
analogy, the \texttt{syscall} interface is a controlled invocation of
kernel services that involve kernel's exclusive rights on sensitive
system operations. In our architecture, \pu holds an exclusive right
to application secrets and sensitive routines with a program, and user
layer must invoke \texttt{privcalls} to enter \pu mode and perform
sensitive operations involving the secrets in a strictly controlled
way. Our architecture allows developers to protect applications secret
within the \pu memory space and also write \privcall routines that
that can securely process the application secret. We developed a
kernel module that adds the support for the \privcall ABI to the Linux
kernel (\lotrkmod). In addition, we provide a library (\liblotr) that
provides the \privcall interface to the user programs and C macros
that enable declaration of \privcall routines, a modified C library
for the building the \pu side (\lotrlibc), and a tool for building
LOTRx86-enabled program (\lotrbuild).

We implemented a prototype of our architecture that is compatible with
\emph{both} Intel and AMD's x86 processors. Based on our prototype, we
developed a proof-of-concept LOTRx86-enabled web server. In our PoC,
the web server's private key is protected in the \pu memory space and
the use of the key (e.g., sign a message with the key) is only allowed
through our \privcall interface. In our PoC web server, in-memory
private key is inaccessible outside the \privcall routines that are
invoked securely, hence an arbitrary access to the key is
automatically thwarted (i.e., HeartBleed). The evaluation of the PoC
and other evaluations are conducted on both Intel and AMD PCs. We
summarize the contributions of our LOTRx86 architecture as the
following:

\begin{itemize}[leftmargin=7mm]
  \setlength\itemsep{0.8em}
\item We propose a portable privileged user mode architecture for
  sensitive application code and data protection that does not require
  address switching or run-time page table manipulation.

\item We introduce the \privcall ABI that allows user layer to invoke
  the \privcall routines in a strictly controlled way. We also provide
  necessary software for building an LOTRx86-enabled software.

\item We developed a PoC LOTRx86-enabled web server to demonstrate the
  protection of in-memory private key during SSL connection.
\end{itemize}

\section{Background: The x86 Privilege Architecture}
\label{sec:background}

The LOTRx86 architecture design leverages the x86 privilege structures
in a unique way. Hence, it is necessary that we explain the x86
privilege system before we go further into the LOTRx86 architecture
design. In this section, we briefly describe the x86 privilege
concepts focusing on the topics that are closely related to this
paper.

\subsection{The Ring Privileges}
\label{sec:ring-privileges}

Modern operating systems on the x86 architecture adapt the two
privilege level model in which user programs run in Ring3 and kernel
in Ring0. The x86 architecture, in fact, supports four privilege
layers -- Ring0 through Ring3 where Ring0 is the highest privilege on
the system. The x86 architecture's definition of privilege is closely
tied to a feature called \emph{segmentation}.

Segmentation divides virtual memory spaces into \emph{segments} which
are defined by a base address, a limit, and a \emph{Descriptor
  Privilege Level (DPL)} that indicates the required privilege level
for accessing the segment. A segment is defined by \emph{segment
  descriptor} in either \emph{Global Descriptor Table (GDT)} or
\emph{Local Descriptor Table (LDT)}. The privilege of an executing
context is defined by a 16-bit data structure called \emph{segment
  selector} loaded in the \emph{code segment register} (\reg{cs}). The
segment selector contains an index to the code segment in the
descriptor table, a bit field to signify which descriptor table it is
referring to (GDT/LDT), and a 2-bit field to represent the
\emph{Current Privilege Level (CPL)}. The CPL in \reg{cs} is
synonymous to the context's current Ring privilege number.

The privilege level (the Ring number) dictates an executing context's
permission to perform sensitive system operations and memory
access. Notably, the execution of privileged instructions is only
allowed to contexts running with Ring0 privilege. Also, the x86 paging
only permits Ring0-2 to access supervisor pages.



\subsection{Memory Protection}
\label{sec:bg-mem-acc}
Operating systems use paging to manage memory access control, and the
segmented memory model has long been an obsolete memory management
technique. However, the paging-based \emph{flat memory model}, which
has become the standard memory management scheme, uses the Ring
privilege levels for page access control. The x86 paging defines
two-page access privilege: User and Supervisor. The Ring 3 can only
access User pages while Ring 0-2 are allowed to access Supervisor
pages\footnote{Intel and AMD have introduced a CPU feature called
  \emph{Supervisor Mode Execution Prevention (SMEP)} and
  \emph{Supervisor Mode Access Prevention (SMAP)}. SMEP prevents
  contexts in Ring 0-2 from executing code in User pages, effectively
  preventing ret2usr style of attacks. SMAP prevents the kernel from
  accessing user pages as data \cite{intel-manual,lwn-smap}}. In
general, the pages in the kernel memory space are mapped as Superuser
such that they are protected from user
applications. \autoref{tbl:ring-priv} outlines the privileges of each
Ring level.

\begin{algorithm}[h]

  \small{
    \begin{algorithmic}[1]

      \Procedure{CG:$R_n$ $\rightarrow$ $R_m$}{$SEGSEL$} \State
      $DESC\_TBL \gets \textbf{if}\ SEGSEL.ti\ ?\ LDT\ :\ GDT$ \State
      $CG \gets DESC\_TBL[SEGSEL.idx]$

      \If {$n$ $>$ $CG.RMPL$ \textbf{or} $n$ $\leq$ $m$} \State
      $\textbf{return}\ DENIED$ \EndIf
   
      \State {Save(\%RIP,\%CS,\%RSP,\%SS)} \Comment {Save caller
        context in temp space}
      \State {$\%SS \gets TSS[m].SS$} \Comment {Load new context to be
        used in Ring $m$} \State {$\%RSP \gets TSS[m].RSP$} \State
      $\%CS \gets CG.TargetCS$ \Comment{Privilege Escalation: $n$
        $\rightarrow$ $m$} \State $\%RIP \gets CG.TargetEntrance$
  
      \State Push\ $SavedSS$ \State Push\ $SavedRSP$ \State Push\
      $SavedCS$ \State Push\ $SavedRIP$ \State RESUME

      \EndProcedure

    \end{algorithmic}

    \caption{x86 callgate operation}
    \label{alg:callgate}
  }
\end{algorithm}
\vspace{-5mm}
\begin{algorithm}[h]
  \small{
  
    \begin{algorithmic}[1]

      \Procedure{Long Return}{} \State \(\triangleright\)
      {\textcolor{darkgray}{can only return to equal or lower
          privileges}} \If {$DestPriv$ $<$ $CurrentPriv$} \State
      $\textbf{return}\ DENIED$ \EndIf \State $\%RIP \gets Pop()$
      \Comment {\textcolor{darkgray}{target addr}} \State
      $\%CS \gets Pop()$ \Comment {\textcolor{darkgray}{target ring
          privilege}} \State $tempRSP \gets Pop()$ \State
      $tempSS \gets Pop()$ \State $\%RSP \gets tempRSP$ \State
      $\%SS \gets tempRSP$ \State RESUME \EndProcedure

    \end{algorithmic}
  
    \caption{x86 long return instruction (\reg{lret}) }
    \label{alg:farret}
  }
\end{algorithm}

\vspace{-5mm}

\subsection{Moving Across Rings}
\label{sec:moving-across-rings}
The x86 architecture provides a number of mechanisms by which a
running context can explicitly invoke privilege escalation for system
services. While the privilege of the context is clearly specified in
its \reg{cs} register, its contents cannot be directly altered (e.g.,
mov \reg{eax}, \reg{cs}) but indirectly with special instructions. The
x86 provides special instructions that allow switching of the code
segment as well as the program counter, namely the \emph{inter-segment
  control transfers} instructions. For instance, The execution of the
\emph{syscall} instruction elevates the \texttt{CPL} of the context to
Ring0 by loading the \reg{cs} with the kernel code segment. It also
loads the PC register (\reg{rip}) with system call entrance point in
the kernel. In modern operating system kernels, only the instructions
that invoke system calls are frequently used. However, it is necessary
that we explain the concepts and mechanisms of the inter-segment
control transfer mechanisms that were introduced along with the four
Ring system long before the instructions dedicated to invoking system
calls.

\textbf{Privilege escalation.} Our design makes use of the
\emph{callgate} mechanism for privilege escalation, a feature present
in all modern (since the introduction of the protected mode) x86
processors. A callgate descriptor can be defined at the descriptor
tables to create an inter-privilege tunnel between the
Rings. Specifically, it defines the target code segment, whose
privilege will be referred to as the \emph{Target Privilege Level
  (TPL)}, a \emph{Target Addr}, and a \emph{Required Minimum Privilege
  Level (RMPL)}.  A context can pass through a call gate via a long
call 
instruction\footnote{"\texttt{lcall}" in AT\&T syntax and
  "\texttt{call\ far}" in Intel syntax} that takes a \emph{segment
  selector} as its operand. The long call instruction first performs
privilege checks when it confirms that the operand given is a
reference to a callgate. A callgate demands its caller's CPL (the
current Ring number) to be numerically equal to or less than (higher
privilege) the callgate's RMPL. Also, the caller's CPL cannot be
numerically less than the TPL of the callgate.  In other words, a
control transfer through a callgate does not allow privilege
de-escalation.  If these privilege checks fail, the context receives a
general protection fault and is forced to terminate. If the privilege
check is successful, the privilege of the context is escalated, and
the program counter (\reg{rip}), as well as the stack pointer
(\reg{rsp}), are loaded with the target address.  A long call
instruction results in privilege escalation if and only if it
references a valid callgate that defines a privilege escalation and
minimum privilege required to enter the callgate.  Therefore a
callgate is a \emph{controlled control transfer} that facilitates
privilege escalation.  We provide a pseudocode that describes the set
of operations performed at the callgate in Algorithm
\autoref{alg:callgate}. Note that we denote a control flow transfer
where a context executing in Ring$n$ enters Ring$m$ through a callgate
using the following notation:

\vspace{1mm} \indent \cg{n}{m}, where $n$ $\leq$ $CG.RMPL$ and $m$
$\leq$ $n$ \vspace{2mm}

\begin{table}[t]
  \centering \small{
    \caption{Privileges of Four Rings on x86}
    \vspace{-3mm}
    \label{tbl:ring-priv}
    \begin{tabular}{|l|l|l|l|l|}
  
      \hline
  
      & Ring0 & \textbf{Ring1} & \textbf{Ring2} & Ring3\\ \hline
      Privileged instruction & $\checkmark$ & $\times$& $\times$ & $\times$ \\ \hline
      Supervisor page access & $\checkmark$ & $\checkmark$ & $\checkmark$ & $\times$ \\ \hline
    
  \end{tabular}
  \vspace{-3mm} }
\end{table}

\textbf{Privilege de-escalation.} A context can return to its original
privilege
mode 
with a long call instruction\footnote{"\texttt{lret}" in AT\&T syntax
  and "\texttt{retf}" in Intel syntax} after privilege escalation. A
long return instruction restores the caller's context that has been
saved by the long call instruction as shown in
Algorithm~\autoref{alg:farret}. It should be noted that a long return
instruction only checks if the destination privilege level is
numerically equal to or greater (lower privilege) by referencing the
saved caller context. In fact, a long return instruction has no way of
knowing if the saved context on the stack is indeed saved by the
callgate. Hence, the long return instruction and similar return
instructions such as \texttt{iret} can be thought of as
\emph{privilege de-escalating control transfer} instructions that pop
the contents that are presumably saved registers.  In this sense, a
long return and its variants provide \emph{non-controlled control
  transfer} mechanism that is used to de-escalate privileges.  We
denote this specific type of control transfer where privilege is
de-escalated (or stays the same) from $m$ to $n$ as the following:

\vspace{1mm} \indent \rtor{m}{n}, where $m$ $\leq$ $n$ \vspace{2mm}

\textbf{Inter-bitness control transfer.} Inter-bitness control
transfer is another type of an x86 control transfer that needs to be
explained before we introduce our design. The x86-64 architecture
provides \emph{32-bit compatibility mode} within the x86-64 (AMD's
amd64 or Intel's IA-32e architecture).  As with the privilege level,
the \emph{bitness} is also defined by the currently active code
segment descriptor.  When a context is executing in a code segment
whose descriptor has the \texttt{L} flag set, the processor operates
in the 64-bit instruction architecture (e.g., registers are 64-bit,
and 64-bit instructions are used).  Otherwise, the context executes as
if the processor is an x86-32 architecture processor.  The bitness
switching, although it changes the processor (current CPU core)
execution mode, is no different than any other inter-segment control
transfers with one exception: a callgate cannot target a 32-bit code
segment.  This is a perk that came with the introduction of the x86-64
implementation.  In summary, we denote 32-bit code segments with a
\emph{x32} suffix as the following:

\vspace{1mm} \indent \rtor{n\_x32}{m} \vspace{2mm}

\section{Attack model and security guarantees}

\subsection{Attack model}

We assume that the adversary is either an outside entity or a
non-administrator user (i.e., no access to root account) who seeks to
extract sensitive application code or data.  The adversary may have an
exploitable vulnerability in the victimized application that could
lead to arbitrary code execution and direct access to application
secret. We assume such vulnerabilities are present when the app has
fully initialized and is servicing its user.  However, we presume that
the program is safe from the adversary during the initialization phase
of the application. We also assume a non-compromised kernel that can
support the LOTRx86 architecture.  Our design requires the presence of
a kernel module that depends on kernel capabilities such as marking
memory regions supervisor or installing custom segment descriptors.
Also, our design includes Enter/Exit gates that facilitate the control
transfer between the \pu and normal user mode.  The gates amount to
about ~50 lines of assembly code and we assume that they are
verifiable and absent of vulnerabilities.

\subsection{Security guarantees}
\label{subsec:sec-gua}
Our work focus on providing developers with an underlying
architecture, a new user privilege layer, which can be leveraged to
protect application secrets and also program routines that access
secrets securely.  Using our architecture, we guarantee that a context
in normal user mode cannot directly access a region protected (as a
part of the \pu memory space) even in the presence of vulnerabilities.
The adversary cannot jump into an arbitrary location in the \pu memory
space to leak secrets since LOTRx86 leverages the x86 privilege
structures to allow only controlled invocation of routines that handle
sensitive information.

On the other hand, we do not focus on the security of the code that
executes in our \pu mode.  We also argue that protection of
application secret in the presence of a vulnerability in the trusted
code base (\pu code in our case) is an unrealistic security objective
for any privilege separation scheme or even hardware-based Trusted
Execution Environments~\cite{arm-trustzone,intel-sgx-web}.  For
instance, a recent work~\cite{jaehyuk} proved that vulnerabilities
inside SGX could be used to disclose protected application secrets.
However, we \emph{do guarantee} that the \pu layer is architecturally
confined to its privilege that it cannot modify kernel memory nor
infringe upon the kernel (Ring0) privileges even in the presence of a
vulnerability in the \pu code.  As we will explain in the coming
section (\autoref{sec:design}), this is a pivotal part of our
architecture design.  The privilege structures and gates that exactly
achieve this security guarantee is one of the key contributions of
this paper.




\section{LOTRx86 Design} \label{sec:design}



\

The primary goal of the LOTRx86 design is to establish a new user
memory protection and access mechanism through the introduction of new
user mode privilege called \pu. Our design eliminates the necessity
for page table switching or manipulation; the access to the protected
memory regions is granted based on the privilege. Our architecture
approaches the problem of application memory safety at a fundamental
level. Instead of leveraging the existing OS-supported protection
mechanisms, we create another privilege distinction within the user
program execution model that resembles the user vs. kernel privilege.

Developers can write sensitive memory handling routines into \pu
layer, then simply place a \privcall in place to invoke a routine that
she defined. Below is the \privcall interface that we provide to
developers:

\begin{code}{C}
privcall(PRIVCALL_NR,...);
\end{code}

The \privcall interface and its ABI is modeled after the Linux
kernel's system call interface. The routine in \pu is identified with
a number (e.g., \texttt{PRIV\_USEPKEY=3}). For developers who have
experience in POSIX system programming, using the \privcalls to
perform operations that involve application secret is intuitive.

\textbf{Privileged-based memory access control.} Our approach
introduces a \emph{privilege-based memory access control}, and it
offers clear advantages over the existing process and thread level
approaches. The cost of the remote procedure calls for bridging two
independent processes, or the cost of page table manipulation is
eliminated. In our architecture, the memory access permissions do not
change when the application secret needs to be accessed. Instead, the
privilege of the the execution mode is elevated to obtain access to
the protected memory.

\textbf{Secure invocation.} \privcall is a single control transfer
instruction (\texttt{lcall}), by which a context enters \pu mode
through the LOTRx86 Enter gate and returns upon finishing the
\privcall routine. Due to this design, the adversary cannot jump into
an arbitrary location with the \pu privilege. Therefore, our
architecture does not experience the security complications inherent
to \emph{enable and disable} models~\cite{shreds}.

\textbf{Portability.} LOTRx86 does not rely on new processor features
for memory
protection~\cite{intel-sgx-web,intel-mpx-web,lwn-mpk,amd-sme}. Instead,
we re-purpose the underused privilege layers to implement \pu. Hence,
our architecture is compatible across all generations of x86-64
processors. As we will present in \autoref{sec:evaluation}, we
evaluated our architecture and a PoC on both Intel and AMD's x86-64
processors.

\textbf{Flexible application privilege separation.}

\begin{figure*}[t!]
  \centering
    
  \captionsetup[subfigure]{width=0.9\textwidth}
  \begin{subfigure}[t]{0.5\textwidth}
    \centering
    \includegraphics[width=1.0\textwidth]{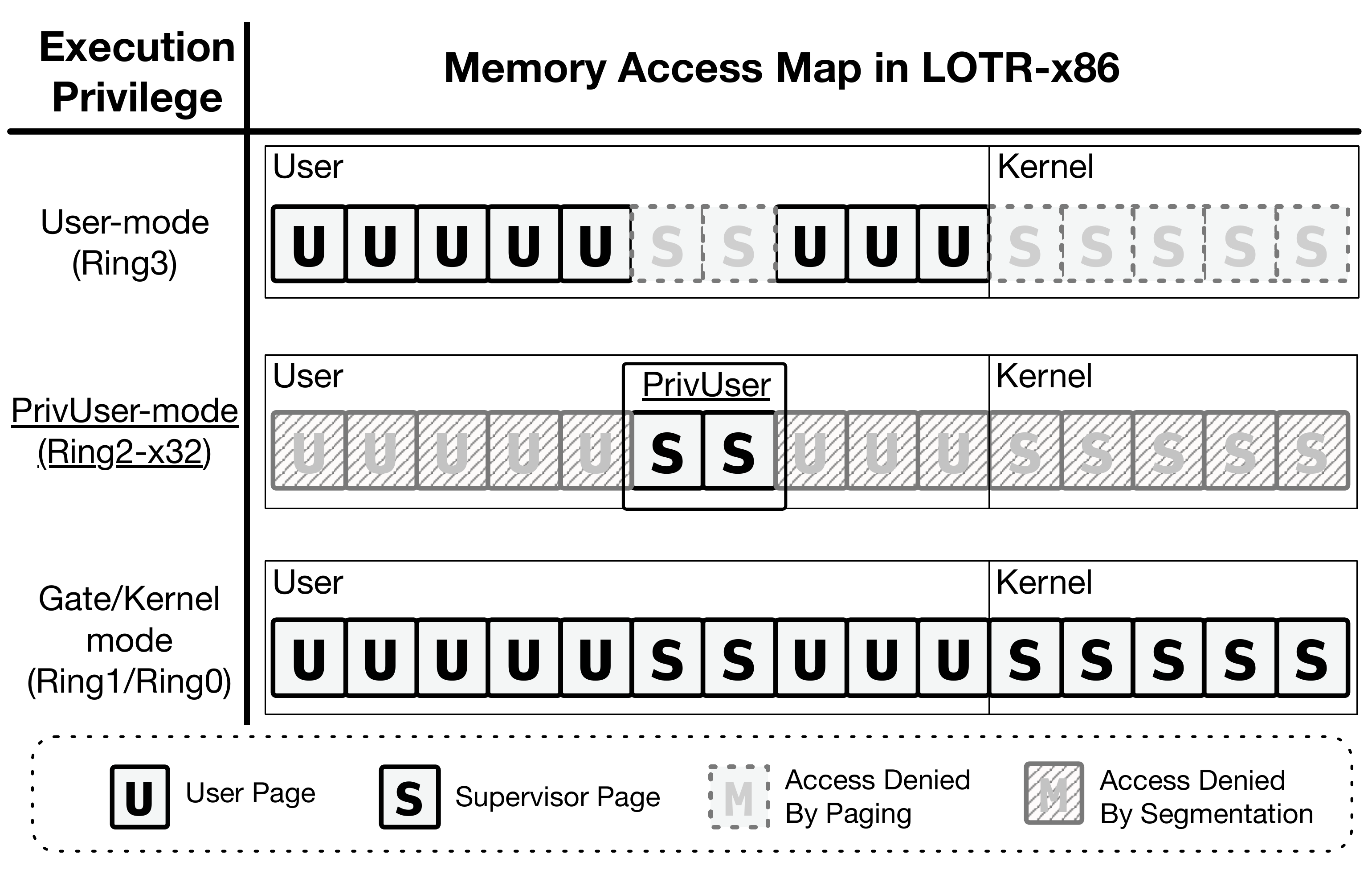}
    \caption{LOTRx86 application memory access map: PrivUser memory
      regions are mapped \emph{Supervisor} protected by paging when in
      User-Mode(Ring3). In PrivUser-Mode (Ring2), in-place memory
      segmentation protects kernel and (optionally) normal user-mode
      memory.}
    \label{fig:mem-map}
  \end{subfigure}
  ~
  \begin{subfigure}[t]{0.5\textwidth}
    \centering \includegraphics[width=1.0\textwidth]{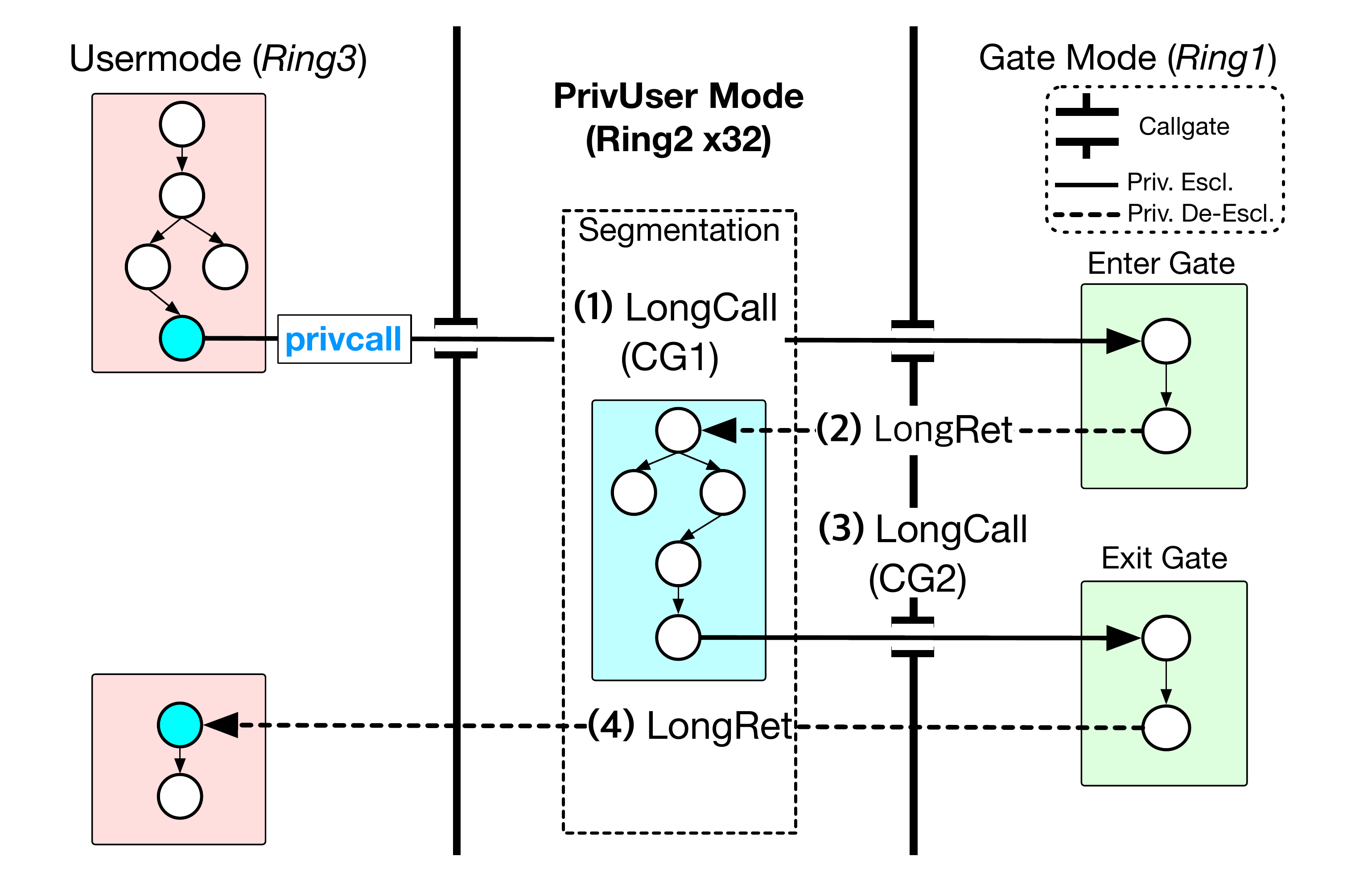}
    \caption{LOTRx86 gate design: implements inescapable segmentation
      enforcement through meticulously designed privilege and gate
      structures. LOTRx86 uses Ring1 as Gate-mode in and out of the
      PrivUser-mode that executes in Ring2-x32.}
    
    \label{fig:gate-design}
  \end{subfigure}
  
  \caption{LOTRx86 architecture overview}
  
  \label{fig:lotr}
  \vspace{5mm}
\end{figure*}

\subsection{Establishing \pu memory space} \label{subsec:harnessing}
We face formidable challenges in the process of establishing the \pu
layer. Our design creates a distinct execution mode (\pu execution
mode) and its address space (\pu address space) for \pu
layer. However, the resulting \pu layer must be intermediate, meaning
that its address space should not be accessible by a user mode
context, and at the same time, \pu execution mode must not be able to
access the kernel address space. However, the x86 paging architecture
provides only two memory privilege distinction: \pgs{U} and
\pgs{S}. The memory segmentation feature that existed in x86-32 is
deprecated in x86-64, eliminating an additional memory access control
mechanism to paging.

In summary, \pu layer must satisfy the two fundamental memory access
security requirements (M-SR1 and M-SR2) to function as an intermediate
layer.

\vspace{2mm} \indent \textbf{M-SR1.} User mode must not be able to
access \pu memory space

\indent \textbf{M-SR2.} \pu mode must not be able to access kernel
memory space \vspace{2mm}

\textbf{Satisfying M-SR1.} We satisfy M-SR1 by mapping all pages that
belong to \pu as \pgs{S} to protect a user mode context from accessing
\pu code and data. As a result, \pu memory space that is mapped as
\pg{S} is accessible to \pu mode, but not to user mode. Now, we see
that we are already using both of the two privilege distinction
recognized by the paging system, and we are unable to protect the
kernel from \pu mode.

\textbf{Solution for M-SR2.} LOTRx86 adapts a scheme that temporarily
enables segmentation when a certain code segment is in use; we enforce
\pu mode to be a \emph{segmentation-enforced execution mode} by
defining it as a 32-bit segmentation-enabled code segment as shown in
\autoref{tbl:lotr-ldt}. This way, entering \pu mode changes not only
the currently active code segment but also the bitness of the
execution mode. That is, when user mode enters \pu mode through
\privcall the execution mode is set to the 32bit compatibility
mode. As a result, we can enforce segmentation to set boundaries to
the powerful \pu mode (Ring2) that are capable of accessing
\pgs{S}. The resulting memory access map of the three execution mode
is illustrated in \autoref{fig:mem-map}. With our design, the \pu
memory space serves as a \emph{functionally} intermediate memory space
for \pu.

\textbf{Remaining challenge.} However, we found that satisfying M-SR2
is a non-trivial issue. The segmentation enforcement alone is not
sufficient for ensuring M-SR2.  As mentioned in
\autoref{subsec:sec-gua}, we must guarantee that the \pu layer has
architecturally well-defined memory boundary against kernel; we must
ensure that all memory access under any circumstances should not be
able to affect kernel.  The challenge is that we must carefully
inspect all possible inter-segment control transfer from \pu mode, to
verify that \pu mode cannot enter a state where it can access kernel
memory.

\subsection{Inescapable segmentation enforcement}
LOTRx86 needs to guarantee that \pu mode is architecturally
confined. Hence, we need to ensure it cannot escape the segmentation
to access kernel memory. More specifically, we need to ensure that no
\emph{non-controlled} (i.e., not through callgates) inter-segment
control transfer paths \emph{out} of the \pu mode arrives in a segment
that is \emph{1.} has a Ring privilege numerically less than 3 (can
access \pgs{S}), \emph{2.} and is a 64-bit segment (no segmentation is
enforced). We denote this control transfer security requirement for
the enforcement of inescapable segmentation as CT-SR, and we explain
how our privilege definitions (i.e., entries in LDT) and our gate
structure (\autoref{fig:gate-design}) satisfy the above requirement.

\vspace{2mm}

\textbf{CT-SR.} \nrtor{2\_x32}{e\_x64} where $e < 3$: there must be no
possible non-controlled control transfer from \pu mode (\ring{2\_x32})
to a 64-bit Ring privilege $e$ that is capable of accessing \pg{S}
access privilege \vspace{2mm}

\textbf{Hardware constraint and gate mode.} Along with the CT-SR,
there is an x86-64 specific perk that has been proved to be a
constraint in our design.  The x86-64 mode (both 64-bit mode and the
32-bit compatibility mode) only supports a 64-bit mode callgate which
is an extended version of its counterpart that existed in
x86-32. Specifically, it does not allow the target code segment of a
callgate to be a 32-bit segment. This implies that an inter-bitness
control transfer through callgate is not supported both ways;
while 
\cg{n\_x32}{m} is possible, but \cg{m}{{n\_x32}} is an invalid
callgate definition. Due to this constraint C, a privilege escalation
and a switch to the 32-bit mode cannot be achieved in a single
callgate transfer. Therefore, we see that we need a separate 64-bit
\gate mode segment to elevate privilege, then enter the 32-bit \pu
mode. However, there exists a more important reason for the existence
of the 64-bit \gate mode and that its privilege \ring{g} must be
higher than that of \pu mode (\ring{p}).

\vspace{2mm}

\indent \textbf{C. \ncg{n}{m\_x32}} : callgate cannot target a 32-bit
code segment

\vspace{2mm}

\textbf{Inspecting non-controlled control transfer routes.} As we
explained in \autoref{sec:background}, an uncontrolled inter-segment
control transfer can be made to jump to a less privileged code segment
without any security checks.  Therefore we must rigorously verify all
possible non-controlled transfers from \ring{p\_x32} to all Ring
levels $e$ that are $e$ $\geq$ $2$ (Ring privilege levels that are
numerically equal or greater, meaning equal or lower privilege). First
of all, we must make sure that a context in \pu mode cannot
arbitrarily jump into an arbitrary place in \gate mode. In order to
prevent a non-controlled control transfer \rtor{p}{g}, we realize that
the gate mode privilege must be higher (numerically lower in terms of
Ring number). Hence the following property must hold in our design:
\vspace{2mm}

\indent \textbf{P1.} $g < p$ (\ring{g} is higher in privilege than
\ring{p}) : privilege of \gate mode must be higher than that of \pu
mode

\vspace{2mm}

The second possible escape route is to perform a same-privilege
inter-bitness (32bit $\rightarrow$ 64bit) inter-segment control
flow. 
We prevent such route by intentionally not defining a 64-bit code
segment for the Ring level 2. A Ring privilege level in the x86
architecture come into existence when it is defined in the descriptor
table, and a context loads the segment selector that points to the
code segment through inter-segment control flow instruction. Hence, a
Ring level that is not defined in the descriptor tables, \emph{does
  not} exist within the system. Hence, by only defining a 32-bit code
segment for Ring2, Ring2 becomes a 32-bit only, segmentation enforced
Ring level in our system definition. We denote this property of our
privilege structure design as the following:

\vspace{2mm} \indent \textbf{P2.} $\nexists$ \ring{p\_x64} : 64-bit
counter part of \pu mode segment must not exist \vspace{2mm}

Our privilege definitions and gate structures (\autoref{tbl:lotr-ldt}
and \autoref{fig:gate-design}) meet the constraint C.  C is satisfied
by the Enter gate in \gate mode.  A \privcall first enters \gate mode
through the CG1 into the Enter Gate (\autoref{tbl:lotr-ldt}).  At the
gate mode, we load the stack with the following arguments: \{\pu entry
point, \pu code segment selector, \pu stack address, \pu stack segment
selector\}, and then perform a far return \texttt{lret} to enter \pu
mode.  While this control transfer is a made through a non-controlled
control transfer instruction, the Enter gate consisting of about ~30
lines of assembly instructions are guaranteed to be executed
\emph{from the beginning} by the CG1. In other words, we chain a
non-controlled control transfer with a controlled control transfer
(CG1) to guarantee its correct execution. Our design also satisfies
CT-SR by maintaining the required properties P1 and P2. We chose Ring1
as the privilege level for \gate mode while enforcing the segmentation
on all \pu mode execution by defining only 32-bit
segmentation-enforced code segment for the Ring level 2. By meeting
CT-SR, we complete our solution for the establishment of the \pu
memory space that satisfies both M-SR1 and M-SR2; the \pu memory space
is protected from context running in the user mode, while \pu mode is
architecturally bound to its memory space that it cannot access kernel
memory under all circumstances.

\section{Prototype Implementation}
\label{sec:implementation}

In this section, we explain the prototype of our LOTRx86 architecture
in detail. Our prototype implementation consists of the following
components:



\begin{itemize}[label={},leftmargin=*]
\item \textbf{lotr-kmod.} We built a Linux kernel module that
  communicates with the host process (LOTRx86 enabled process). The
  module creates a virtual device interface at \texttt{/dev/lotr}, and
  an LOTRx86 enabled program communicates with our kernel module with
  the \texttt{ioctl} interface. The kernel module builds the \pu-space
  for the program when requested.
  
\item \textbf{liblotr.} The user library \texttt{liblotr} allows
  developers to the use of our architecture in the host program,
  isolate the application secrets, and implement \privcalls that
  securely access the secrets. A developer can initialize the
  \pu-space and utilize the \privcall interface through our user
  library. The library also includes tools and scripts for building
  the executable that runs in the \pu-space.
  
\item \textbf{lotr-libc.} We provide a modified version of the
  \emph{musl}~\cite{musl} libc for building the \pu executable. We
  modified the heap memory manager such that only \texttt{S-pages} are
  allocated to the heap managers used in the \pu mode. In this way, we
  prevent the leakage of the application secret and the by-products of
  its processing to the user space.
\item \textbf{lotr-build.} \texttt{lotr-build} is a collection of
  compilation scripts and tools that help developers in compiling the
  \pu portion of their application and incorporating it into the host
  application. We further explain this procedure later.
\end{itemize}

\subsection{PrivUser mode Initialization}

The \lotrkmod kernel module initializes the LOTRx86 infrastructure
such as the \gate-mode, \pu mode and control transfer structures for
the host process. The host application is required to call
\texttt{init\_lotr(\&req)} function from \liblotr with an argument of
the \texttt{struct init\_request} type during its initialization. The
request structure contains the addresses and sizes of \pu components
that \lotrkmod need in its initialization routine. Such information
includes the range of \pu code segment, data segment, the entry point
for the \pu-space, pages to be used as a stack in \pu, and so
forth. The addresses of the segments are available through the symbols
generated by our build tools during the compile-time, while the stack
is allocated through \texttt{mmap} in \liblotr. Additionally,
\lotrkmod contains the Enter gate and Exit gate that are loaded into
the kernel memory upon module load.

The \lotrkmod kernel module creates an LDT for the host process and
writes the segment and callgate descriptors that are used for the
\gate-mode and \pu mode. Unlike the GDT, an LDT is referenced on a
\emph{per-process} basis; an LDT can be created for each process, and
the register that points to the currently active LDT called
\emph{ldtr} is updated in each context switch. For this reason, the
LOTRx86 descriptors can only be referenced by the host process that
explicitly requested the initialization of the LOTRx86
infrastructure. \lotrkmod creates the descriptor segments listed in
\autoref{tbl:lotr-ldt}. A set of Ring1 code and data segments are used
for the \gate-mode, and Ring2-32bit segment descriptors are loaded as
a context enters the \pu mode.

The initialization also set the \gate-mode stack to be loaded at the
Enter callgate. As briefly explained in \autoref{sec:background}, the
x86 callgate mechanism finds the address of the new stack for the
control transfer at the TSS structure. The TSS structure holds the
addresses of for each Ring levels. In our case, we use two callgates,
CG($R_3$$\rightarrow$$R_1$) and CG($R_{2\_x32}$$\rightarrow$$R_1$),
that both require a stack for Ring1. Hence, we allocate stack space
and record the top of the stack in the Ring1 stack field of the TSS
(TSS.SP1).

Another important task carried out during the initialization (in
\lotrkmod) is marking the pages that belong to the \pu-space
Supervisor pages. The kernel module walks the page tables and marks
\pu pages Supervisor by clearing the \texttt{User} bit in the page
table entry. All pages that are marked Supervisors are maintained in a
linked list so they can be reverted or freed when necessary as the
host process terminates.

When all necessary initialization procedures are finished, the kernel
module creates a lock for the host process based on its PID. From this
point on, \lotrkmod ignores additional initialization request
delivered via the \texttt{ioctl} requests from the host to thwart any
possible attempt to compromise the \pu-space.


\begin{table}[t]
  \centering

  \small{

    \begin{tabular}{|l|l|l|}

      \hline
      & Type                  & Priv.        \\ \hline
      Gate-mode CS     & Code Segment          & Ring1           \\ \hline
      Gate-mode DS     & Data Segment          & Ring1           \\ \hline
      PrivUser mode CS & Code Segment          & Ring2-x32       \\ \hline
      PrivUser mode DS & Data Segment          & Ring2-x32       \\ \hline
      CG1   & CG1(R3$\rightarrow$R1) & CPL $\leq$ 3 \\ \hline
      CG2    & CG2(R2$\rightarrow$R1) & CPL $\leq$ 2 \\ \hline

\end{tabular}
\vspace{2mm}
\caption{LOTRx86 LDT descriptors: by defining segment and callgate
  descriptors in LDT, LOTRx86 creates \gate-mode and \pu mode for a
  process}
\label{tbl:lotr-ldt}
} \vspace{-8mm}
\end{table}

\subsection{LOTRx86 ABI}

The \privcall interface of the LOTRx86 is almost identical to the
\texttt{syscall} interface; \privcall follows the x86-64 System V
AMD64 ABI system call convention~\cite{amd64-abi}. That is, we use
\reg{rax}, \reg{rdi}, \reg{rsi}, \reg{rdx}, \reg{r10}, \reg{r8},
\reg{r9} registers for passing arguments to the \pu mode, and the
return value is stored in \reg{rax}. Underneath the surface, however,
our unique design enables establishment and secure use of the
\pu-space. From here on, we explain each stage of the control flow
transfers in the ABI -- starting from a \privcall and its return to
its caller.

\textbf{\privcall interface.} A \privcall(NR\_PRIVCALL, ...) consists
of layers of macros that handle a variable number of arguments and
place them in the argument registers in order. After the arguments are
placed according to the x86-64 syscall ABI, a long call
(\texttt{lcall}) instruction is executed with a segment selector that
points to the Enter callgate as an argument. Upon the executing of
\texttt{lcall}, the execution continues at the Enter gate with a
privilege of Ring1.


% ...

% ...

\begin{figure}[!t] \centering
\begin{asm}{GAS}
# Entered from privcall in user mode
^\texttt{\textbf{LOTREnterGate:}}^
# (a) Allow only Ring 3 to enter this gate
movq 8(%rsp), %r11
cmp $3, %r11
jnz EXIT
# (b) Save User mode(R3) Context
pushq 24(%rsp);
pushq 16(%rsp);
pushq 8(%rsp);
pushq 0(%rsp);
SAVE_REGS();
# (c) Transfer Arguments into PrivUser Stack 
movq $PrivUserStack, %r11;
subq $60, %r11;
movl $DummyEIP, 0(%r11d);
movq %rax, 4(%r11d);
movq %rdi, 12(%r11d);
movq %rsi, 20(%r11d);
movq %rdx, 28(%r11d);
movq %r10, 36(%r11d);
movq %r8, 44(%r11d);
movq %r9, 52(%r11d);
# (d) Push PrivUser(RIP,CS,RSP,SS) onto stack,
# then perform control flow transfer
movq $PrivUserEnter, %r9;
pushq $PrivUserSS;
pushq %r11;
pushq $PrivUserCS;
pushq %r9;
lret;
# Entered from privret in PrivUser mode
^\texttt{\textbf{LOTRExitGate:}}^
sub $GateContextSize, %rsp
RESTORE_REGS();
# in case security check (a) fails
EXIT:
lret;
\end{asm}
  \caption{Simplified pseudo assembly code of LOTRx86 Enter gates}
  \label{fig:gate-code}
  \vspace{-5mm}
\end{figure}

\textbf{Enter gate.} The LOTRx86 Enter gate plays a pivotal role in
safeguarding the user mode context that invoked a \privcall into the
\pu mode. \autoref{fig:gate-code} is simplified pseudo assembly code
of the implementation. The Enter gate is written in assembly code and
is about ~30 instructions that carry out three main operations.

First, the Enter gate checks the saved \reg{cs} in the gate stack. At
this point, the ring privilege has been escalated to that of the \gate
mode (Ring1), stack pointer now points to \gate mode stack, and the
caller context is saved in the new stack.  (for detailed x86 callgate
operation, revisit Algorithm \autoref{alg:callgate} in
\autoref{sec:background}). The least significant 2 bits of the saved
\reg{cs} (\reg{cs}[1:0]) indicates the caller's Ring privilege.  By
ensuring the value to be 3, we prevent \pu mode from entering the
Enter gate for possibly malicious intent.

Then the gate saves the user mode caller context in the \gate mode
stack. Note that the x86 long call instruction has context saving
feature built in. However, since we use the Ring1 for both Enter gate
and Exit gate, the saved context is overwritten when the context
returns from \pu mode back to the Exit gate. Therefore, we found that
it is necessary to perform a manual context saving of the four
registers (\reg{RIP}, \reg{CS}, \reg{RSP}, \reg{SS}) in the beginning
of our Enter gate as shown in the code block $(b)$ in
\autoref{fig:gate-code}.

The second operation (code block $(c)$ in \autoref{fig:gate-code})
illustrates the transforming of the \privcall arguments that follow
the x86-64 calling convention into that of the \pu mode ABI; the
in-register arguments must be transferred to the \pu mode stack as
preparation before entering the \pu mode. Unlike the conventional
x86-32 ABI, we use the 64-bit arguments in the \pu mode by
default. The fact that the \pu mode runs in 32-bit compatibility mode
but uses 64-bit length arguments is a peculiar characteristic of our
design, and the LOTRx86 Enter gate resolves the calling convention
discrepancy.

The last operation (code block $(d)$) performed in at the Enter gate
is to transfer the control flow into the entry point of \pu mode. We
push the entry point address, the address of the \pu mode stack that
contains the arguments passed on by the \privcall in the user mode at
this point, and their segments (\reg{cs} and \reg{ss}) on to the
current (\gate mode stack). Then, we execute the \texttt{lret}
instruction to enter the \pu mode.

\textbf{PrivUser entry point.} The \pu mode entry point first performs
a bound check on the \reg{eax} that contains the \privcall number
(i.e., $1$ $\leq$ \texttt{nr\_Privcall} $\leq$ $MAX\_PRIVCALL$). The
pointers to the predefined \privcall routines are arranged in the
\emph{Privuser Call Table (PCT)} whose role is identical to the
\emph{system call table} in the Linux kernel. This mechanism prevents
a maliciously crafted \privcall from calling an arbitrary memory
address. If the check is valid then the entry point calls the
\emph{wrapper function} for the \privcall routine that corresponds to
the number is invoked.

\textbf{PrivUser routine.} The developers can define a \privcall
routine through the \texttt{PRIVCALL\_DEFINE(func\_name,...)} macro.
The macro creates and exports a wrapper function that calls the main
function. This particular implementation is borrowed from the Linux
kernel~\cite{kernel-web}. The wrapper casts the 64-bit arguments into
function-specific argument sizes (e.g., 64-bit to int (32bit)) for the
defined \privcall routine. After the \privcall routine is finished,
the execution returns to the \pu entry point to be concluded by
\texttt{lcall} that transfers the control flow \emph{back} into the
Exit gate with the privilege of the \gate mode (Ring1).

\textbf{Exit gate.} The exit gate scrubs the scratch registers (the
six general purpose registers as stated in the System V i386 calling
convention) to prevent information leakage from the \pu mode. Recall
that we manually saved the user mode context in the stack from the
Enter gate. We subtract the stack pointer ($48 (8\times 6)$ bytes in
our implementation) to move it to the saved context. We execute
\texttt{popq} instruction to restore \reg{rbp} then the \texttt{lret}
instruction to restore \reg{RIP}, \reg{CS}, \reg{RSP}, \reg{SS} to
return to the original caller of the \privcall with a privilege of
user mode (Ring3).


\subsection{Developing LOTRx86-enabled program}

\begin{figure*}[t]
  \centering \includegraphics[width=0.95\textwidth]{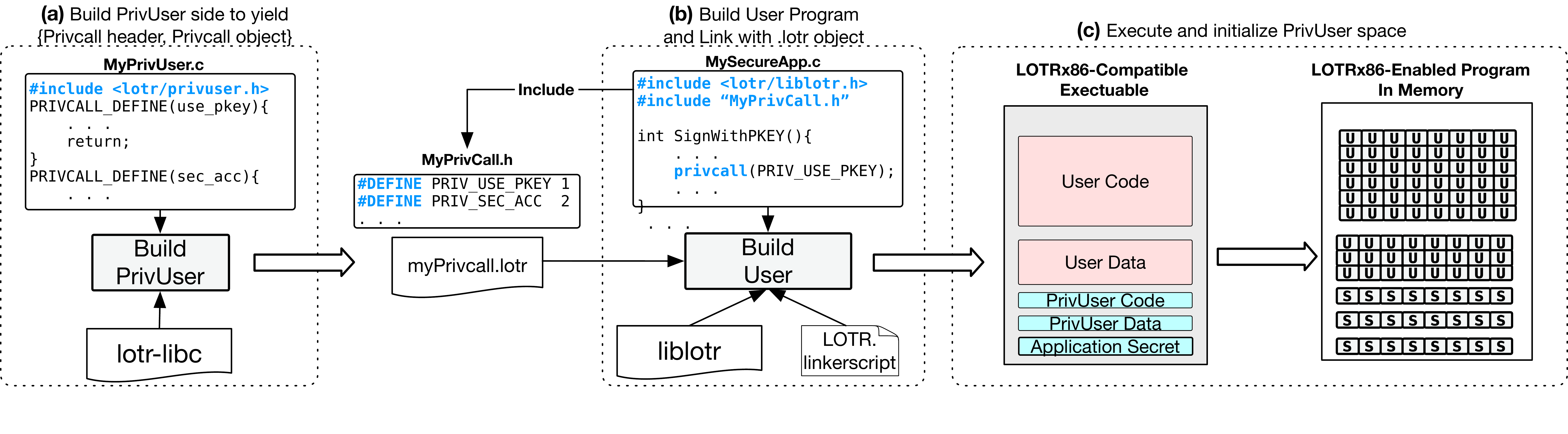}
  \vspace{-8mm}
  \caption{Building LOTRx86 compatible executable}
  \label{fig:lotr-build}
  \vspace{5mm}
\end{figure*}

We developed tools and libraries that allow developers to write
LOTRx86-enabled program. Writing a \privcall routine is similar to
writing a regular user-level code. However, there are a few key
differences both in developer's perspective and underneath the
surface. Here, we outline the important aspects in LOTRx86-enabled
program development. \autoref{fig:lotr-build} illustrates the overall
build process of a LOTRx86-enabled executable.

The \privcall interface and the development of \privcall routines are
intentionally modeled after the Linux kernel's system call interface.
For this reason, the procedures for developing the \pu side of the
program and invoking them as necessary are nearly identical to those
of developing new system calls to the kernel.

\textbf{\privcall declaration.} \liblotr provides two important macros
through \texttt{<lotr/privuser.h>}.  First is the declaration macro
\#PRIVCALL\_DEFINE.  The macro takes the name of the function as the
first argument and up to six arguments.  The type and the name of the
arguments must be entered as if they are separate argument (e.g.,
(int, mynumber)).  This is because \texttt{PRIVCALL\_DEFINE} generates
a wrapper function that casts the ABI-defined arguments into the
argument's type.  We restrain from further explaining the details of
the macro since it is almost identical to the kernel's
\texttt{SYSCALL\_DEFINE} macro.

\textbf{Compiling with \lotrmusl.} We provide \texttt{gcc-lotr} which
is a wrapper to the \texttt{gcc} compiler.  \texttt{gcc-lotr} links
the user's \pu code with \lotrlibc instead of the default glibc
(32bit).  \lotrlibc is a modified version of musl-libc.  We modified
the \texttt{malloc} function in the musl-libc so that it manages a
memory block from the \pu memory space \pgs{S}.  This is to prevent
the by-products or the application itself from being placed in a
memory region accessible to the normal user mode.  Additionally, we
implemented a function that initializes process \emph{Thread Local
  Storage (TLS)} that can be called from \liblotr's
\texttt{init\_lotr()} function.  The initialization of a process TLS
is performed by the libc library before the program's \texttt{main()}
is executed.  Therefore, it is necessary to implement a separate
function to initialize the TLS for LOTRx86.

\textbf{Argument passing.}  \liblotr provides an argument page that is
always allocated within the 32-bit address space.  Developers can
first copy the argument into the argument page then pass the reference
to \pu mode.  Alternatively, developers can use \texttt{LOTR\_SECRET}
keyword to global variables to force them to be placed in a data
section called \texttt{.lotr\_secret} which is loaded into the \pu
memory space.

\textbf{Building final exectuable.} Compiling the \pu code with our
build tools yields two files: a header file in which \privcall numbers
are defined, and a LOTRx86 object in \texttt{.lotr} extension.  The
header file lists the assigned number for each declared \privcall
routines, and the \texttt{.lotr} file is an object file ready to be
linked to the main program.  Our build tool compiles the \pu code in
x86-32 code.  However, we copy the sections of the 32-bit object into
a new 64-bit ELF object format so that it can be linked into the main
program.  The \pu build tools also strip all the symbols to prevent
symbol collision between the 32-bit libc (\lotrmusl) and the 64-bit
libc used in the main program, then it generates a symbol table that
includes addresses of the \pu object sections and most importantly,
the \pu entry point.  The main program is built with our linker script
(\texttt{LOTR.linkerscript}) that loads the symbol table generated
during \pu build.  When the main program launches,
\texttt{init\_lotr()} fetches the symbols and transfers them to
\lotrkmod, and the kernel module marks the memory pages that belong to
the \pu memory space \pgs{S}.

\subsection{Kernel changes}
The LOTRx86 prototype is implemented as a kernel module. However, we
also made minor but necessary modifications to the Linux kernel. First
of all, we made sure that system calls (e.g., \texttt{mprotect}) that
alter the memory permissions of the user memory space ignore the
request when the affected region includes \pu-memory. This is achieved
by simply placing a ``\emph{if-then-return -ERR}'' statement for the
case where the address belongs to the user-space but the page is an
\texttt{S-page}. We made a similar change to the \texttt{munlock}
system call so that \pu's \texttt{P-pages} are excluded from possible
memory swap-outs.

\section{Proof-of-Concepts and Evaluation}
\label{sec:evaluation}
To show the feasibility and efficiency of the LOTRx86 architecture
approach, we develop a proof-of-concept (PoC) by incorporating our
architecture into the \emph{Nginx} web server~\cite{nginx} as well as
the \emph{LibreSSL}~\cite{libressl} that is used by the web browser to
support SSL. We modified the parts of the web server to protect the
in-memory private key in the \pu memory space and only allow accesses
to the key through our \privcall interface. In this section, we
present the results from a set of microbenchmarks that we performed to
measure the latency induced by a \privcall. Then we compare the
performance of the PoC web server whose private key is protected with
its original version. Our experiments are conducted on both Intel and
AMD to show that our approach to show that our approach is
portable. It should be noted that the PoC and all examples are
compiled as \emph{64-bit programs}. The specifications of the two x86
machines are as follows:

\begin{itemize}[label={},leftmargin=*]
\item \textbf{Intel-based PC:} i7-4770 @ 3.40GHz, 4 cores, 16GB RAM
\item \textbf{AMD-based PC:} Ryzen 7 1800X @ 3.60GHZ, 8 cores, 32GB
  RAM
\item \textbf{OS on both PC:} Ubuntu 16.04 LTS, kernel version 4.13
\end{itemize}

\subsection{Microbenchmarks}

The \privcall allows developers to invoke routines that access
application secrets in the \pu layer. A certain amount of added
latency is inevitable since we perform a chain of control transfers to
securely invoke the \privcall routines. In this experiment, we compare
the latency of an empty \privcall against the commonly used library
calls to show that the added overhead is indeed a reasonable trade-off
for the protection of application secrets. We also conducted the
microbenchmark in two varying setups. In the first setup, we built
executables that makes a single invocation of each calls (\privcalls,
library calls), and we produced the results by executing the
executables 1000 times. In the second setup, we measure the latency of
a 1000 consecutive invocations of each calls using a for loop. These
two setups represent the two situations where \privcall is
infrequently called and frequently called.

\begin{figure*}[t!]
  \centering

  \captionsetup[subfigure]{width=0.9\textwidth}
  \begin{subfigure}[t]{0.5\textwidth}
    \centering
    \includegraphics[width=0.95\textwidth]{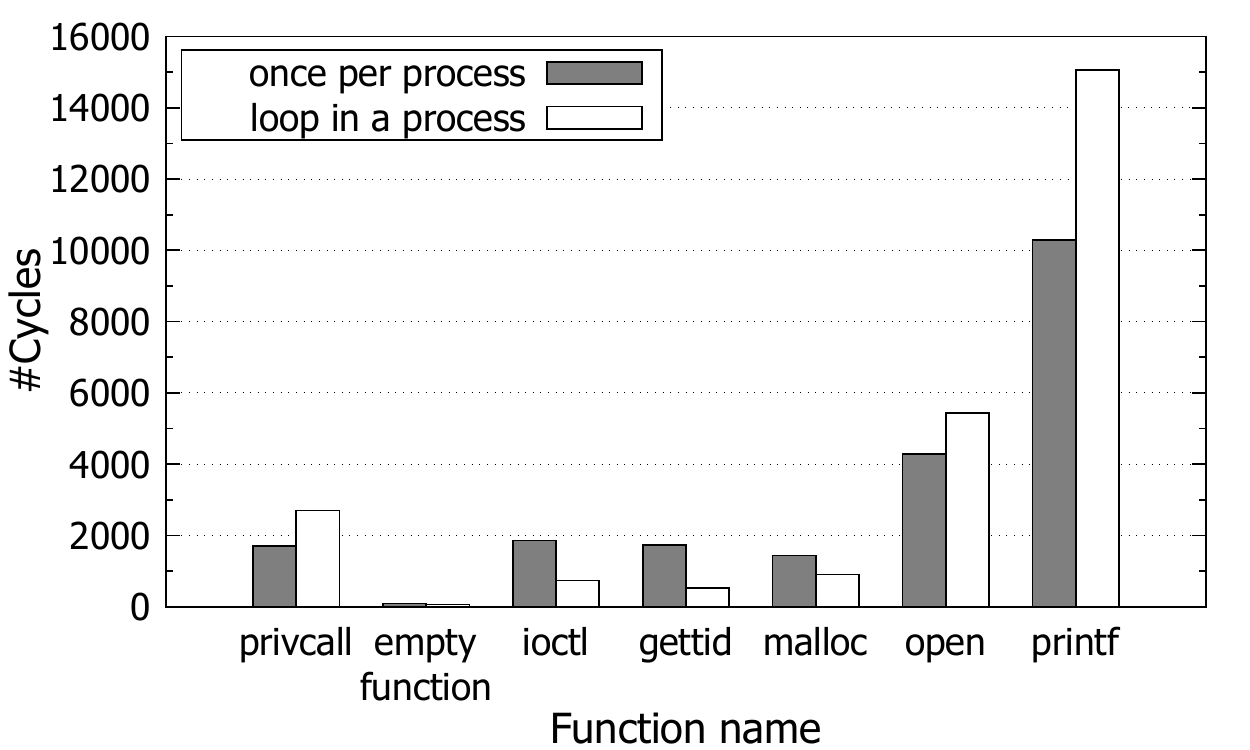}
    \caption{Micro-benchmark (Intel): privcall vs. common C library
      calls}
    \label{fig:microbench_intel}
  \end{subfigure}
  ~
  \begin{subfigure}[t]{0.5\textwidth}
    \centering
    \includegraphics[width=0.95\textwidth]{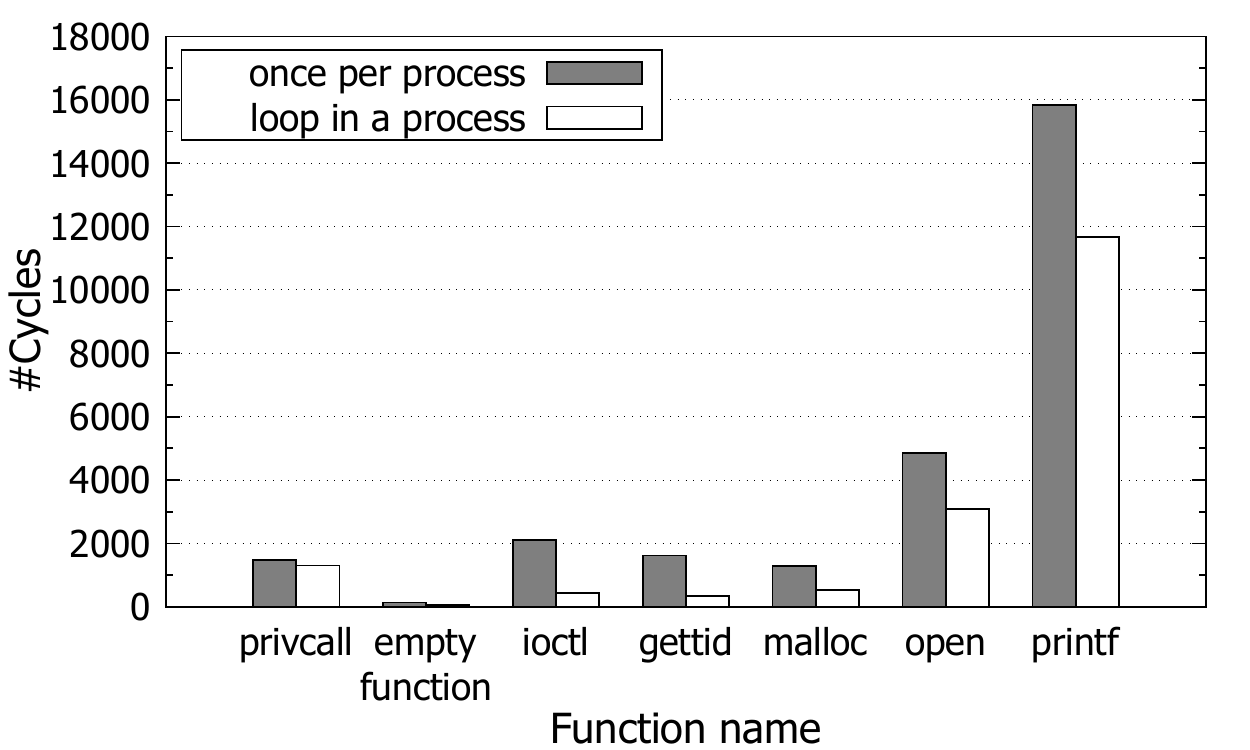}
    \caption{Micro-benchmark (AMD): privcall vs. common C library
      calls}
    \label{fig:microbench_amd}
  \end{subfigure}

  \begin{subfigure}[t]{1\columnwidth}
    \centering
    \includegraphics[width=0.95\textwidth]{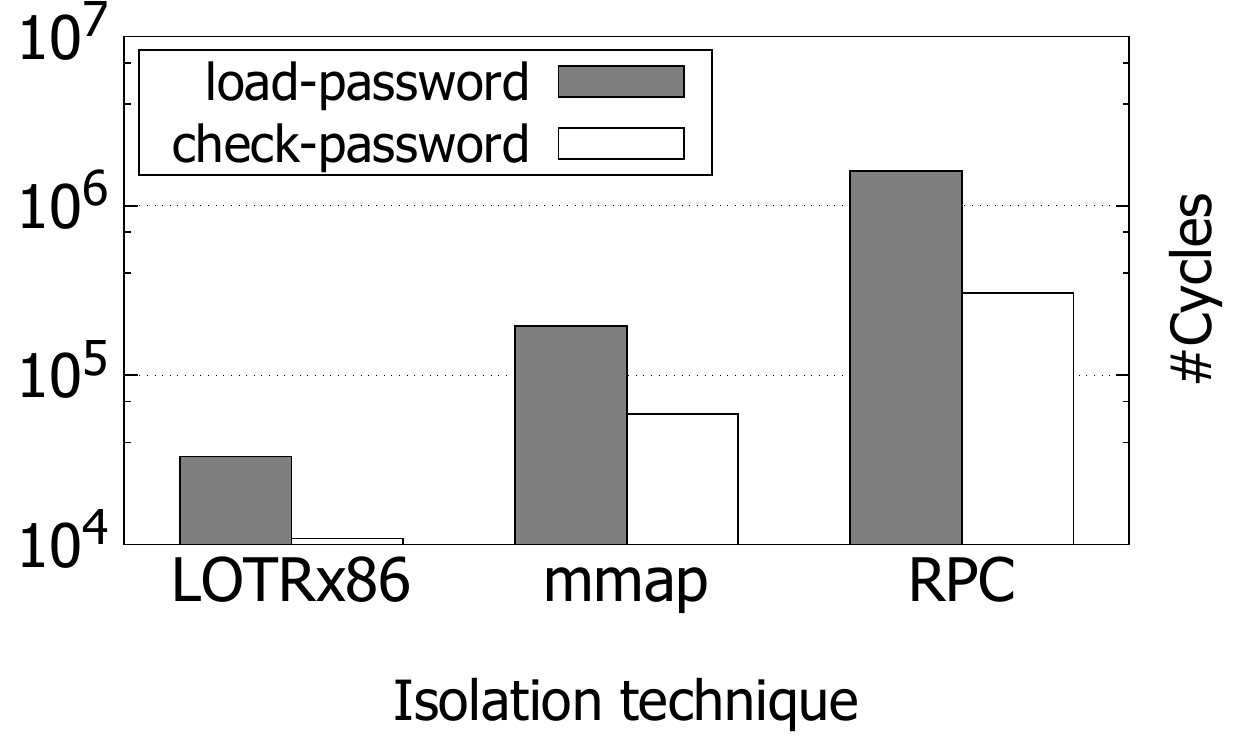}
    \caption{Execution time of LOTRx86 vs. traditional memory
      protection methods (Intel)}
    \label{fig:microbench2_intel}
  \end{subfigure}
  ~
  \begin{subfigure}[t]{1\columnwidth}
    \centering
    \includegraphics[width=0.95\textwidth]{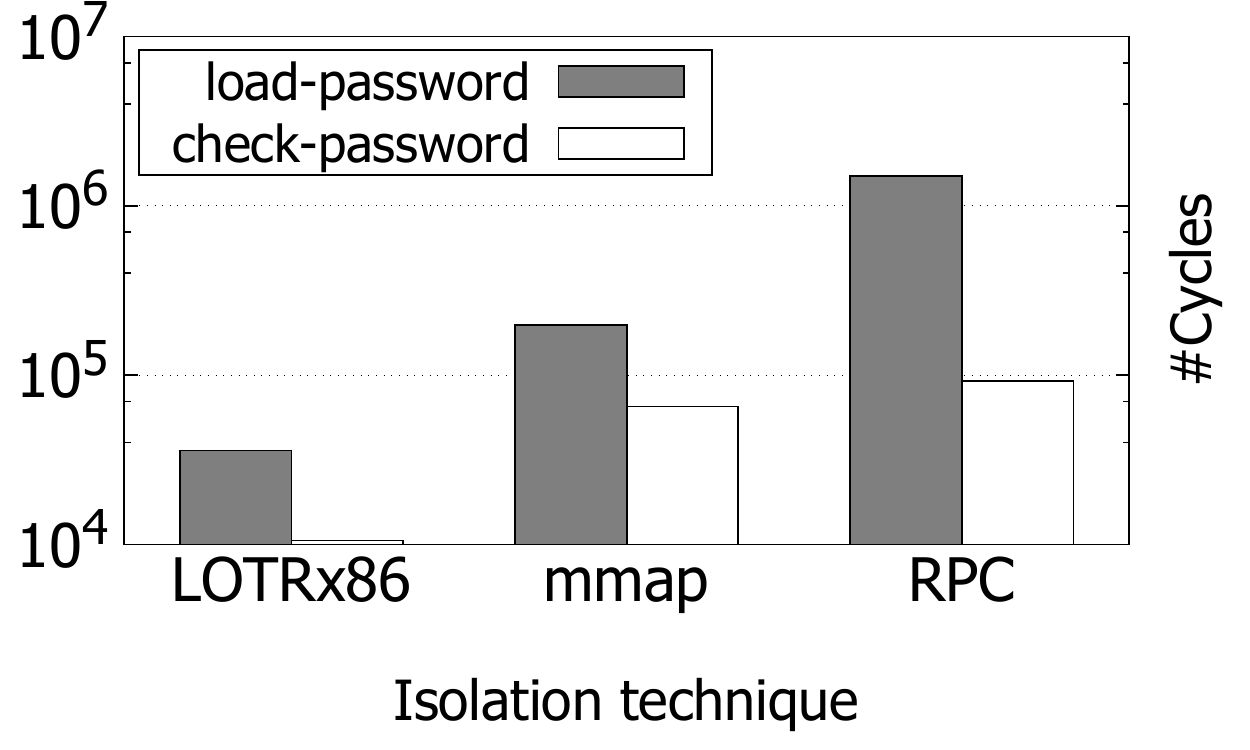}
    \caption{Execution time of LOTRx86 vs. traditional memory
      protection methods (AMD)}
    \label{fig:microbench2_amd}
  \end{subfigure}
  \caption{Micro-benchmark \privcall vs common library calls (a,b) and
    comparison against traditional memory protection methods (c,d)}
 
  \label{fig:microbench2}
  \vspace{1mm}
\end{figure*}

~\autoref{fig:microbench_intel} shows the experiment results on the
Intel PC while ~\autoref{fig:microbench_amd} shows the results from
the same experiment on the AMD PC. The latency incurred by a \privcall
proves to be at a reasonable level. The single invocation performance
is on par with the most basic library calls such as \texttt{ioctl} or
\texttt{gettid}, consuming about 1500-2000 cycles on both PCs. It is
noticeable that the latency of a \privcall does not improve
drastically as some of the other calls such as \texttt{gettid} whose
number of cycles has dropped from 2148 to 404 on Intel PC. As to this
result, we surmise that the control flow transfer chain used in our
architecture affect the caching behavior of the processor
negatively. Also, the libc and kernel's system call invocation have
been extremely well optimized for a long period of time. Hence, we
plan to investigate possible optimizations that can be applied to
LOTRx86 in future. However, LOTRx86 is the only portable solution on
the x86 architecture that achieves in-process memory protection. Also
it is the most efficient solution among portable solutions. We present
a comparison of LOTRx86 against the traditional memory protection
techniques to support our claim. Moreover, we argue that the
performance overhead of LOTRx86 is at a reasonable level in an
application scale. We present a macro benchmark results using our PoC
(Nginx with LOTRx86).




\subsection{Comparison with traditional memory protection techniques}
We implemented a simple demonstration in-process memory protection
using LOTRx86, page table manipulation technique implemented with
\texttt{mmap} and \texttt{mprotect}, and a socket-based remote
procedure call mechniasm (from \texttt{<rpc/rpc.h>}).

\textbf{Test program.} Our simple program first load a password from a
file into the protected memory region, then it receives an input from
user via stdin to compare it against the protected password. In more
detail, we implemented two functions \texttt{load\_password} and
\texttt{check\_password} using the three protection mechanisms to
evalulate their performance overhead. For page table based method, we
use \texttt{mprotect} to set the page that contains the
\texttt{load\_password} and \texttt{check\_password} and the page
dedicated for storing the loaded password to \texttt{PROT\_NONE}. In
case of the RPC mechanism we simply place the two measured functions
and the password-storing buffer in a different process and make RPCs
to execute the functions remotely.

\textbf{Performance overhead comparison.} The measurements for the
execution time of the two functions, implemented with three different
mechanisms, are illustrated in \autoref{fig:microbench2}. We averaged
the results from 1000 trials (the y-axis is in log scale). The results
show that LOTRx86 proves to be much faster than the two traditional
methods by a large margin. On Intel PC, LOTRx86 greatly reduces the
execution time (33051 cycles) of \texttt{load\_password} by 83.11\%
(195661 cycles) and by 97.93\% (1600291 cycles), compared to
\texttt{mmap}- and RPC-based implementation,
respectively. \texttt{check\_password}. This is because LOTRx86 does
not require page table modifications that may cause system-wide
performance overhead, nor the cost of communication with an external
entity.


\begin{figure*}[t!]
  \centering
    
  \captionsetup[subfigure]{width=0.9\textwidth}
  \begin{subfigure}[t]{0.5\textwidth}
    \centering \includegraphics[width=0.95\textwidth]{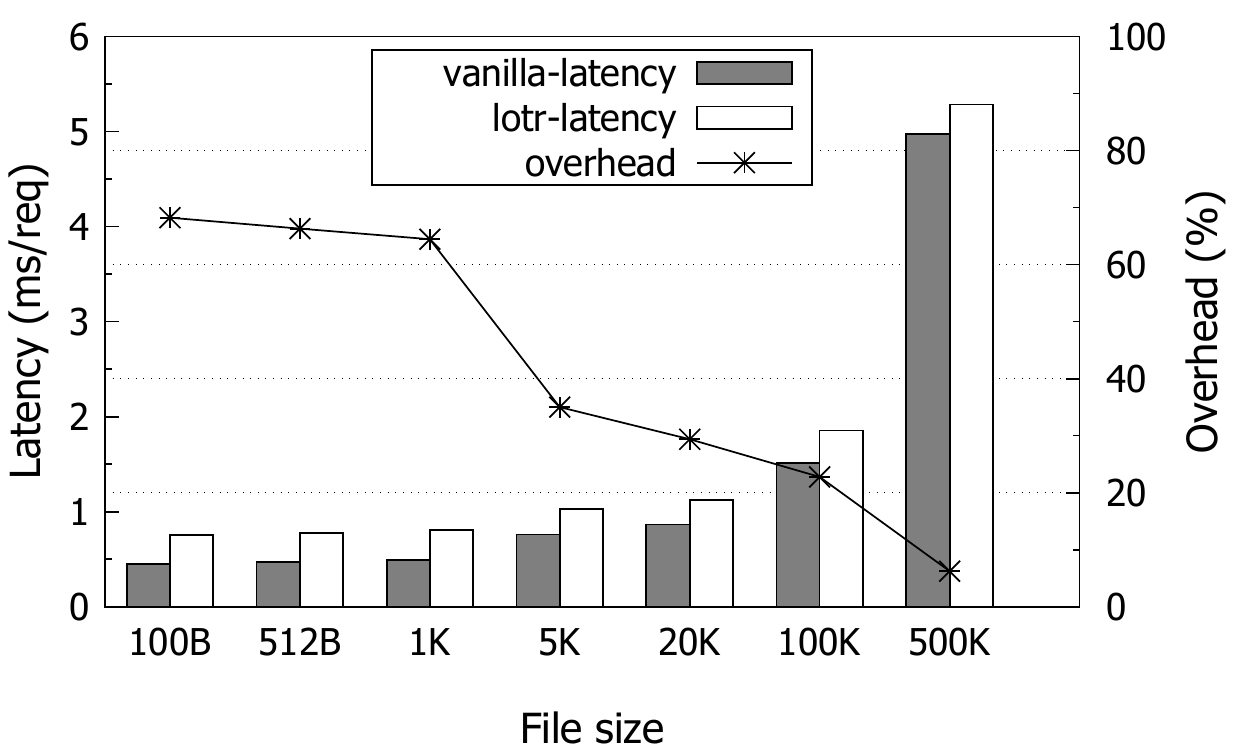}
    \caption{Nginx latency measurements on Intel}
    \label{fig:nginx_intel}
  \end{subfigure}
  ~
  \begin{subfigure}[t]{0.5\textwidth}
    \centering \includegraphics[width=0.95\textwidth]{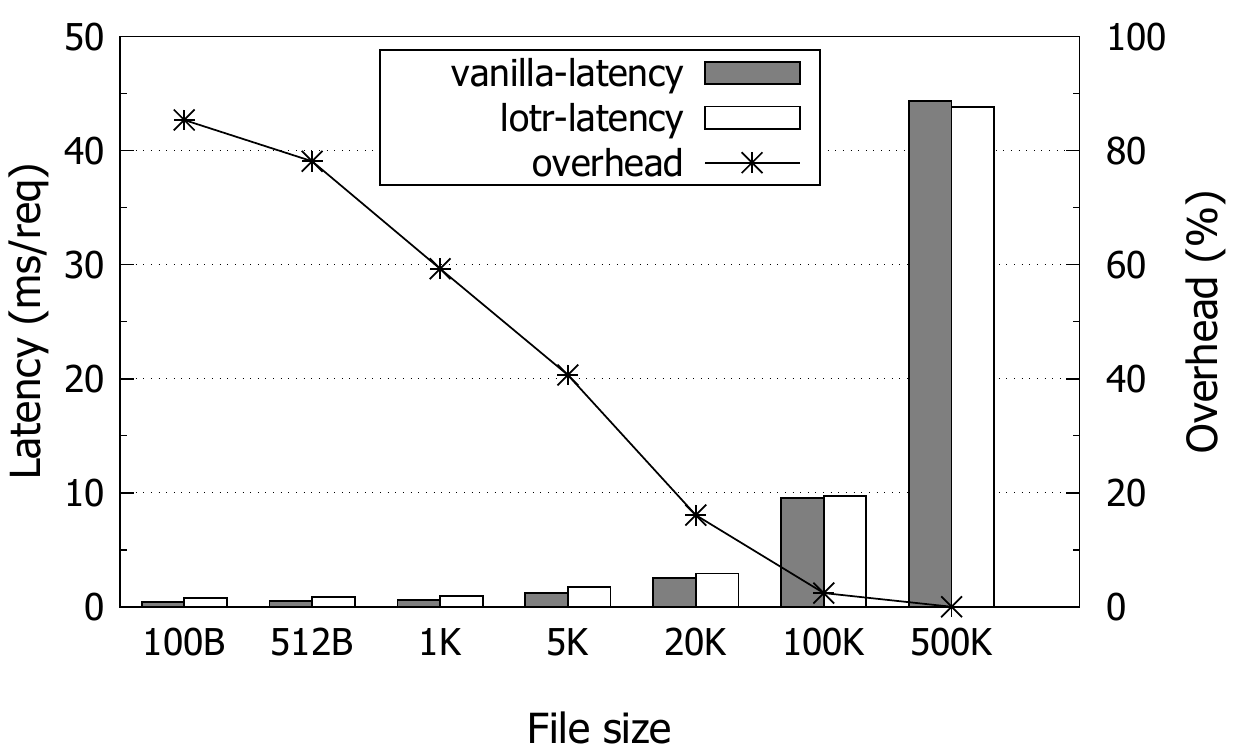}
    \caption{Nginx latency measurements on AMD}
    \label{fig:nginx_amd}
  \end{subfigure}
  
  \label{fig:microbench}
  \caption{SSL KeepAlive response latency with varying file sizes on
    LOTRx86-enabled Nginx}
  \vspace{5mm}
\end{figure*}
\subsection{LOTRx86-enabled web server}
To develop a proof-of-concept LOTRx86-enabled web server, we made
changes to LibreSSL and the Nginx web browser. Specifically, we
replaced parts of the software that accesses private keys with a
\privcall routine that performs the equivalent task. In the resulting
web server's process address space, the private key always resides in
the \pu memory space. Therefore, any arbitrary memory access (e.g.,
buffer over-read in HeartBleed) are thwarted. Only through the
pre-defined \privcall routines, the web server can perform operations
that involve the private key.

\textbf{Implementation.} During its initialization, Nginx loads the
private key through a function called
\texttt{SSL\_CTX\_use\_PrivateKey\_file}. This function performs a
series of operations to read the private key then parsing the contents
into a ASN1 structure, then the function eventually produces an RSA
structure that is used by LibreSSL during SSL connections.  We
re-implemented the function using \privcalls. In our version of the
function, the opening of the file and loading its contents into memory
are performed in \pu mode the structures that contain the private key
or its processed forms, are stored in the \pu memory space.  For
passing arguments, we created a custom C structure that contains the
necessary information that needs to be passed via \privcalls.  Once
the private key is converted into a RSA data structure, it is stored
safely in the \pu memory space until it needs to be accessed during
the handshake stage in an SSL connection.  During the handshake,
server digitally signs a message using the private key to authenticate
itself to its client. We modified the \texttt{RSA\_sign()} such that
it makes \privcalls to request operations involving the RSA structure.
In more detail, we copy the message to be signed in the argument page
shared between user mode and \pu mode that is designated by \liblotr.


\textbf{Performance measurements.} We used the \textit{ab} apache
benchmark tool to perform a benchmark similar to the one performed in
~\cite{secage}, a work that leverages hypervisor to achieve a similar
objective to LOTRx86. Using the tool, we make 1000 KeepAlive requests
to the server, then the server responds by sending a file back to the
client.  In the benchmark, we measured the average execution time from
the socket connection to the last response from the server; The we
varied the size of the requested file size to represent different
configurations (we used \{1k, 5k, 20k, 50k, 100k, 500k\} and
~\cite{secage} uses \{5k,20k,50k\}. The results are shown in
\autoref{fig:nginx_intel} and \autoref{fig:nginx_amd}.  Note that due
to the difference of CPU performance, the range of y-axes is
different.

The additional performance overhead due to LOTRx86 mainly comes from
the execution mode transition (user mode to \pu mode). A total of
three \privcall invocations are made in opening and loading the
contents of the private key file into a buffer in \pu memory space,
and a single \privcall to sign the message using the private key. In
case of 5K requested file size, a rather extreme case, LOTRx86 adds
about $35$\% on Intel processor and $40$\% on AMD. However, the
overhead becomes relatively irrelevant as the file size increases:
20K: $29.36$\% (Intel), 16.07\% (AMD), 500K : 6.25\% (Intel) ~0\%
(AMD). This particular experiment tests the feasibility of LOTRx86 in
latency-critical tasks, and the results show that our approach is
feasible even in such cases.

\vspace{-1mm}

\section{Related Work}

The LOTRx86 architecture proposes a method for construction of a
privileged userspace is capable of protecting designated set of user
application code and data. The new architecture involves the underused
intermediate rings and segmentation-based memory bounds. In this
section we discuss previous work on user-space memory protection and
system privilege restructuring methods for system fortification. Our
architecture creates a virtual intermediate memory privilege only
using the existing features in the x86 architecture. There have been a
number of approaches that adapted varying memory protection
techniques.

\textbf{Alternative Privilege Models.} Nested
kernel~\cite{nestedkernel} introduced a concept of inner-kernel that
takes control of the hardware MMU by deprivileging the original kernel
by disabling a subset of its Ring0 power. Nested Kernel exports
virtual MMU interface that allows the deprivileged kernel to request
sensitive memory management operation explicitly.

The x86-32 hypervisor implementation before the introduction of
Intel's hardware-assisted virtualization features such as Intel's VT-x
and AMD's SVM~\cite{intel-manual,amd-svm-manual}, made use of the
intermediate Rings and segmentation to achieve
virtualization. Hypervisor
implementations~\cite{xen-paper,virtualbox,vmware-paper} deprivileged
the operating system kernel by making them run in the intermediate
Rings then enforced segmentation to protect the in-memory
hypervisor. LOTRx86 not only explores the use of the intermediate
Rings on 64-bit operating systems but also for a very different
purpose. Our architecture inhabits the abandoned Rings into more
privileged user mode in which developers can place their sensitive
application code and data.

\textbf{Use of processor features.} The segmentation feature present
in x86-32 had been employed in a number of approaches for application
memory protection~\cite{nacl,vx32}(and as aforementioned in early
hypervisor implementations). Both Native Client(NaCl) and Vx32
provides a safe is a user-level sandbox that enables safe execution of
guest plug-ins to the host program. In this regard, the attack model
of these work differ from that of LOTRx86. LOTRx86 assumes that the
host application is untrusted and we place sensitive code and data in
the PrivUser space. Also, The Nacl sandbox has adopted SFI to
compensate for the lack of segmentation in x86-64~\cite{nacl-ext}.
LOTRx86 enables inescapable segmentation enforcement to construct \pu
mode in x86-64.

Processor architectures have been extended to support user-space
memory protection. Intel has recently introduced Software Guard
eXtensions (SGX) that creates an enclave which a predefined set of
code and data can be protected~\cite{intel-sgx-web}. Intel also ships
memory bound checking functionality called MPX~\cite{intel-mpx-web},
which provides hardware assist for bound checking that software fault
isolation approaches advocated. Intel has also disclosed plans for
domain-based memory partitioning that resembles the memory domain
feature in the ARM architecture. More recently AMD has disclosed a
white paper describing the upcoming memory encryption feature to be
added to its x86 processors~\cite{amd-sme}.

The fragmented hardware support for application-level memory
protection served as a pivotal motivation for our approach. Our design
does not rely on any specific hardware feature and preserves
portability. However, there is a difference in attack model. SGX
distrusts kernel and the protected user memory within its enclave
stays intact even under kernel compromise. On the other hand, LOTRx86
is incapable of operating in a trustworthy way when the kernel is
compromised. Nevertheless, we argue that our work presents an unique
approach that achieves in-process memory protection while preserving
portability.

\textbf{Process/thread level partitioning.} Early work on application
privilege separation focused on restructuring a program into separate
processes~\cite{ppe,privman,privtrans,nacl}. In essence, placing
program components into process-level partitions aims to achieve
complete address space separation.The approach presents a number of
disadvantages. First, the approach inevitably involves a
\emph{Inter-Process Communication (IPC)} mechanism to establish a
communication channel between the partitions so that they can remotely
invoke functions (i.e., \emph{Remote Procedure Calls (RPC)}) in other
partitions and pass arguments as necessary.

More recent work used threads as a unit of separation compartment that
prevents leakage of sensitive
memory~\cite{wedge,salus,arbiter,smv}. Chen et al. ~\cite{shreds}
pointed out that even the thread compartments are still too
coarse-grained, introduce a high performance overhead due to page
table switches, and requires developers to make structural changes to
their program. Their work Shreds takes advantage of the memory domain
feature on the ARM architecture to create a secure code block within
the program. However, the Domain-based memory partitioning is only
available has been deprecated on \texttt{AARCH64}.

Unlike the apporaches that retrofit the OS protection mechanisms, our
approach makes a fundamental change in privilege architecture. Another
significant difference is that our approach does not require address
space switch nor run-time page table modifications. In our design, the
\emph{privilege} is what changes when granted access to the protected
memory through \privcall; the page tables that map the protected
memory region as \texttt{S-pages} are intact whether the running
context is in the normal user mode or in \pu mode.

\textbf{Software-based fault isolation (SFI).} Software-based fault
isolation
techniques~\cite{sfi,xfi,nacl,nacl-ext,sfi-cisc,vx32,rocksalt} employ
software techniques such as compilers or instrumentation to create
logical fault domains within the same address space often to contain
code execution or memory access. SFI is often used to partition an
untrusted module into a sandbox to protect the host
program~\cite{nacl,nacl-ext,vx32,rocksalt}. The attack surface that
LOTRx86 covers is the exact opposite, or inverse, of their approach;
our work seeks to protect a sensitive and more privileged user code
and data from rest of the program.

\textbf{Hypervisor-based Approaches.}
A number of works have leveraged hypervisors to protect applications
in virtualized systems.
memory~\cite{secage,inktag,overshadow,minibox,trustvisor,yang,sego}. Hypervisor-based
approaches leverage hypervisor-controlled page tables and other
hypervisor control over the virtualized system to ensure
trustworthiness of applications and system services. Similar to SGX,
hypervisor-based approaches are designed on the premise that kernel is
vulnerable or possibly malicious. Additionally, these works assume the
presence of a hypervisor on the system.

\section{Limitations and future work}
LOTRx86 proposes a novel approach to application memory
protection. However, the architecture is still at its infancy. We
describe the limitations of the current prototype and discuss issues
that needs to be addressed.

\textbf{Argument passing.} While the LOTRx86 supports passing of
64-bit arguments to \pu, \pu mode is confined to a 32-bit address
space. Hence, \pu mode is incapable of accessing 64-bit pointers to
further reference the members of an object. However, passing a complex
object with multiple pointer members to a trusted execution mode is
generally discouraged~\cite{intel-sgx-arg}. Such practices will expose
the trusted execution mode to a large attack surface. We advise
developers to clearly define the necessary arguments with known fixed
sizes for a \privcall routine.

\textbf{Further optimization.} We believe there is a room for further
optimizations for the LOTRx86 architecture. However, finding
resourceful optimization guides for using the intermediate Rings were
absent due to its rare usage in modern operating systems. We plan to
investigate further to find ways to improve the performance of our
architecture.

\textbf{SMEP/SMAP.} Intel's SMEP and SMAP~\cite{intel-manual} prevents
supervisor mode (Ring0-2) to access or execute \pgs{U}. SMEP does not
affect LOTRx86 since \pu does not execute any code in
\pgs{u}. However, SMAP prevents \pu mode from accessing the argument
page shared with user mode. One possible solution is to implement a
system call or an \texttt{ioctl} call that toggles the SMAP
enforcement such that \pu mode can fetch data from the shared
page. Note that kernel's \texttt{copy\_from\_user} API also
temporarily disables SMAP to copy from user-supplied pointer to the
argument.

\textbf{ASLR in \pu memory space.} We currently build \pu executable
statically, then copy its sections into a 64-bit object that can be
linked to the main program executable. While the only a small and
easily verifiable amount of code should reside in \pu in principle,
ASLR support in \pu memory space is one of future work.

\section{Conclusion}
We presented LOTRx86, a novel approach that establishes a new user
privilege layer called \emph{PrivUser} that protects and safeguards
secure access to application secrets. The new \pu memory space is
protected from user mode access. We introduced the \privcall interface
that provides user mode a controlled invocation mechanism of the \pu
routines to securely perform operations involving application
secrets. Our design introduced a unique privilege and control transfer
structures that establish a new user mode privilege. We also explained
how our design satisfies the security requirements for the \pu layer
to have a distinct execution mode and memory space. In our evaluation,
we showed that the latency added by a \privcall is on par with
frequently used C function calls such as \texttt{ioctl} and
\texttt{malloc}. We also implemented and evaluated the LOTRx86-enabled
Nginx web server that securely accesses its private key through the
\privcall interface. Using the Apache \emph{ab} server bench mark
tool, we measured the average keep-alive response time of the server
to find the average overhead incurred by LOTRx86 in various response
file size. The average overhead is limited to $30.40$\% on the Intel
processor and $20.19$\% on the AMD processor.

\bibliographystyle{ACM-Reference-Format} \balance
\bibliography{lotr-arxiv.bib}


\begin{thebibliography}{00}


\ifx \showCODEN    \undefined \def \showCODEN     #1{\unskip}     \fi
\ifx \showDOI      \undefined \def \showDOI       #1{#1}\fi
\ifx \showISBNx    \undefined \def \showISBNx     #1{\unskip}     \fi
\ifx \showISBNxiii \undefined \def \showISBNxiii  #1{\unskip}     \fi
\ifx \showISSN     \undefined \def \showISSN      #1{\unskip}     \fi
\ifx \showLCCN     \undefined \def \showLCCN      #1{\unskip}     \fi
\ifx \shownote     \undefined \def \shownote      #1{#1}          \fi
\ifx \showarticletitle \undefined \def \showarticletitle #1{#1}   \fi
\ifx \showURL      \undefined \def \showURL       {\relax}        \fi
\providecommand\bibfield[2]{#2}
\providecommand\bibinfo[2]{#2}
\providecommand\natexlab[1]{#1}
\providecommand\showeprint[2][]{arXiv:#2}

\bibitem[\protect\citeauthoryear{??}{mus}{2018}]%
        {musl}
 \bibinfo{year}{2018}\natexlab{}.
\newblock \bibinfo{title}{musl libc}.
\newblock \bibinfo{howpublished}{\url{https://www.musl-libc.org}}.
  (\bibinfo{year}{2018}).
\newblock
\newblock
\shownote{Last accessed Jan 23, 2018.}


\bibitem[\protect\citeauthoryear{AMD}{AMD}{2013}]%
        {amd-svm-manual}
AMD \bibinfo{year}{2013}\natexlab{}.
\newblock \bibinfo{booktitle}{{\em AMD64 Architecture Programmer's Manual}}.
\newblock AMD.
\newblock


\bibitem[\protect\citeauthoryear{Barham, Dragovic, Fraser, Hand, Harris, Ho,
  Neugebauer, Pratt, and Warfield}{Barham et~al\mbox{.}}{2003}]%
        {xen-paper}
\bibfield{author}{\bibinfo{person}{Paul Barham}, \bibinfo{person}{Boris
  Dragovic}, \bibinfo{person}{Keir Fraser}, \bibinfo{person}{Steven Hand},
  \bibinfo{person}{Tim Harris}, \bibinfo{person}{Alex Ho},
  \bibinfo{person}{Rolf Neugebauer}, \bibinfo{person}{Ian Pratt}, {and}
  \bibinfo{person}{Andrew Warfield}.} \bibinfo{year}{2003}\natexlab{}.
\newblock \showarticletitle{Xen and the art of virtualization}. In
  \bibinfo{booktitle}{{\em Proceedings of the nineteenth ACM symposium on
  Operating systems principles}} {\em (\bibinfo{series}{SOSP '03})}.
  \bibinfo{publisher}{ACM}, \bibinfo{address}{New York, NY, USA},
  \bibinfo{pages}{164--177}.
\newblock
\showISBNx{1-58113-757-5}
\showDOI{%
\url{https://doi.org/10.1145/945445.945462}}


\bibitem[\protect\citeauthoryear{Baumann, Peinado, and Hunt}{Baumann
  et~al\mbox{.}}{2015}]%
        {haven}
\bibfield{author}{\bibinfo{person}{Andrew Baumann}, \bibinfo{person}{Marcus
  Peinado}, {and} \bibinfo{person}{Galen Hunt}.}
  \bibinfo{year}{2015}\natexlab{}.
\newblock \showarticletitle{Shielding Applications from an Untrusted Cloud with
  Haven}.
\newblock \bibinfo{journal}{{\em ACM Trans. Comput. Syst.\/}}
  \bibinfo{volume}{33}, \bibinfo{number}{3}, Article \bibinfo{articleno}{8}
  (\bibinfo{date}{Aug.} \bibinfo{year}{2015}), \bibinfo{numpages}{26}~pages.
\newblock
\showISSN{0734-2071}
\showDOI{%
\url{https://doi.org/10.1145/2799647}}


\bibitem[\protect\citeauthoryear{Bittau, Marchenko, Handley, and Karp}{Bittau
  et~al\mbox{.}}{2008}]%
        {wedge}
\bibfield{author}{\bibinfo{person}{Andrea Bittau}, \bibinfo{person}{Petr
  Marchenko}, \bibinfo{person}{Mark Handley}, {and} \bibinfo{person}{Brad
  Karp}.} \bibinfo{year}{2008}\natexlab{}.
\newblock \showarticletitle{Wedge: Splitting Applications into
  Reduced-privilege Compartments}. In \bibinfo{booktitle}{{\em Proceedings of
  the 5th USENIX Symposium on Networked Systems Design and Implementation}}
  {\em (\bibinfo{series}{NSDI'08})}. \bibinfo{publisher}{USENIX Association},
  \bibinfo{address}{Berkeley, CA, USA}, \bibinfo{pages}{309--322}.
\newblock
\showISBNx{111-999-5555-22-1}
\showURL{%
\url{http://dl.acm.org/citation.cfm?id=1387589.1387611}}


\bibitem[\protect\citeauthoryear{Brumley and Song}{Brumley and Song}{2004}]%
        {privtrans}
\bibfield{author}{\bibinfo{person}{David Brumley} {and} \bibinfo{person}{Dawn
  Song}.} \bibinfo{year}{2004}\natexlab{}.
\newblock \showarticletitle{Privtrans: Automatically Partitioning Programs for
  Privilege Separation}. In \bibinfo{booktitle}{{\em Proceedings of the 13th
  Conference on USENIX Security Symposium - Volume 13}} {\em
  (\bibinfo{series}{SSYM'04})}. \bibinfo{publisher}{USENIX Association},
  \bibinfo{address}{Berkeley, CA, USA}, \bibinfo{pages}{5--5}.
\newblock
\showURL{%
\url{http://dl.acm.org/citation.cfm?id=1251375.1251380}}


\bibitem[\protect\citeauthoryear{Bugnion, Devine, Rosenblum, Sugerman, and
  Wang}{Bugnion et~al\mbox{.}}{2012}]%
        {vmware-paper}
\bibfield{author}{\bibinfo{person}{Edouard Bugnion}, \bibinfo{person}{Scott
  Devine}, \bibinfo{person}{Mendel Rosenblum}, \bibinfo{person}{Jeremy
  Sugerman}, {and} \bibinfo{person}{Edward~Y. Wang}.}
  \bibinfo{year}{2012}\natexlab{}.
\newblock \showarticletitle{Bringing Virtualization to the x86 Architecture
  with the Original VMware Workstation}.
\newblock \bibinfo{journal}{{\em ACM Trans. Comput. Syst.\/}}
  \bibinfo{volume}{30}, \bibinfo{number}{4}, Article \bibinfo{articleno}{12}
  (\bibinfo{date}{Nov.} \bibinfo{year}{2012}), \bibinfo{numpages}{51}~pages.
\newblock
\showISSN{0734-2071}
\showDOI{%
\url{https://doi.org/10.1145/2382553.2382554}}


\bibitem[\protect\citeauthoryear{Chen, Garfinkel, Lewis, Subrahmanyam,
  Waldspurger, Boneh, Dwoskin, and Ports}{Chen et~al\mbox{.}}{2008}]%
        {overshadow}
\bibfield{author}{\bibinfo{person}{Xiaoxin Chen}, \bibinfo{person}{Tal
  Garfinkel}, \bibinfo{person}{E.~Christopher Lewis}, \bibinfo{person}{Pratap
  Subrahmanyam}, \bibinfo{person}{Carl~A. Waldspurger}, \bibinfo{person}{Dan
  Boneh}, \bibinfo{person}{Jeffrey Dwoskin}, {and} \bibinfo{person}{Dan~R.K.
  Ports}.} \bibinfo{year}{2008}\natexlab{}.
\newblock \showarticletitle{Overshadow: A Virtualization-based Approach to
  Retrofitting Protection in Commodity Operating Systems}.
\newblock \bibinfo{journal}{{\em SIGPLAN Not.\/}} \bibinfo{volume}{43},
  \bibinfo{number}{3} (\bibinfo{date}{March} \bibinfo{year}{2008}),
  \bibinfo{pages}{2--13}.
\newblock
\showISSN{0362-1340}
\showDOI{%
\url{https://doi.org/10.1145/1353536.1346284}}


\bibitem[\protect\citeauthoryear{Chen, Reymondjohnson, Sun, and Lu}{Chen
  et~al\mbox{.}}{2016}]%
        {shreds}
\bibfield{author}{\bibinfo{person}{Y. Chen}, \bibinfo{person}{S.
  Reymondjohnson}, \bibinfo{person}{Z. Sun}, {and} \bibinfo{person}{L. Lu}.}
  \bibinfo{year}{2016}\natexlab{}.
\newblock \showarticletitle{Shreds: Fine-Grained Execution Units with Private
  Memory}. In \bibinfo{booktitle}{{\em 2016 IEEE Symposium on Security and
  Privacy (SP)}}. \bibinfo{pages}{56--71}.
\newblock
\showDOI{%
\url{https://doi.org/10.1109/SP.2016.12}}


\bibitem[\protect\citeauthoryear{Corbet}{Corbet}{2012}]%
        {lwn-smap}
\bibfield{author}{\bibinfo{person}{Jonathan Corbet}.}
  \bibinfo{year}{2012}\natexlab{}.
\newblock \bibinfo{title}{Supervisor mode access prevention}.
\newblock \bibinfo{howpublished}{\url{https://lwn.net/Articles/517475/}}.
  (\bibinfo{year}{2012}).
\newblock


\bibitem[\protect\citeauthoryear{Corbet}{Corbet}{2015}]%
        {lwn-mpk}
\bibfield{author}{\bibinfo{person}{Jonathan Corbet}.}
  \bibinfo{year}{2015}\natexlab{}.
\newblock \bibinfo{title}{Memory protection keys}.
\newblock \bibinfo{howpublished}{\url{https://lwn.net/Articles/643797/}}.
  (\bibinfo{year}{2015}).
\newblock


\bibitem[\protect\citeauthoryear{Corperation}{Corperation}{2018a}]%
        {intel-sgx-web}
\bibfield{author}{\bibinfo{person}{Intel Corperation}.}
  \bibinfo{year}{2018}\natexlab{a}.
\newblock \bibinfo{title}{Intel\textsuperscript{\textregistered} Software Guard
  Extensions (Intel SGX)}.
\newblock \bibinfo{howpublished}{\url{https://software.intel.com/en-us/sgx}}.
  (\bibinfo{year}{2018}).
\newblock
\newblock
\shownote{Last accessed Feb 27 , 2018,.}


\bibitem[\protect\citeauthoryear{Corperation}{Corperation}{2018b}]%
        {intel-mpx-web}
\bibfield{author}{\bibinfo{person}{Intel Corperation}.}
  \bibinfo{year}{2018}\natexlab{b}.
\newblock \bibinfo{title}{Introduction to
  Intel\textsuperscript{\textregistered} Memory Protection Extensions}.
\newblock
  \bibinfo{howpublished}{\url{https://software.intel.com/en-us/articles/introduction-to-intel-memory-protection-extensions}}.
    (\bibinfo{year}{2018}).
\newblock
\newblock
\shownote{Last accessed Feb 22 , 2018,.}


\bibitem[\protect\citeauthoryear{Corperation}{Corperation}{2018c}]%
        {amd64-abi}
\bibfield{author}{\bibinfo{person}{Intel Corperation}.}
  \bibinfo{year}{2018}\natexlab{c}.
\newblock \bibinfo{title}{System V Application Binary Interface}.
\newblock
  \bibinfo{howpublished}{\url{https://software.intel.com/sites/default/files/article/402129/mpx-linux64-abi.pdf}}.
    (\bibinfo{year}{2018}).
\newblock
\newblock
\shownote{Last accessed Feb 21 , 2018,.}


\bibitem[\protect\citeauthoryear{Corporation}{Corporation}{2017}]%
        {virtualbox}
\bibfield{author}{\bibinfo{person}{Oracle Corporation}.}
  \bibinfo{year}{2017}\natexlab{}.
\newblock \bibinfo{title}{VirtualBox Technical documentation}.
\newblock
  \bibinfo{howpublished}{\url{https://www.virtualbox.org/wiki/Technical\_documentation}}.
    (\bibinfo{year}{2017}).
\newblock
\newblock
\shownote{Last accessed Aug 23, 2017.}


\bibitem[\protect\citeauthoryear{Dautenhahn, Kasampalis, Dietz, Criswell, and
  Adve}{Dautenhahn et~al\mbox{.}}{2015}]%
        {nestedkernel}
\bibfield{author}{\bibinfo{person}{Nathan Dautenhahn},
  \bibinfo{person}{Theodoros Kasampalis}, \bibinfo{person}{Will Dietz},
  \bibinfo{person}{John Criswell}, {and} \bibinfo{person}{Vikram Adve}.}
  \bibinfo{year}{2015}\natexlab{}.
\newblock \showarticletitle{Nested Kernel: An Operating System Architecture for
  Intra-Kernel Privilege Separation}.
\newblock \bibinfo{journal}{{\em SIGARCH Comput. Archit. News\/}}
  \bibinfo{volume}{43}, \bibinfo{number}{1} (\bibinfo{date}{March}
  \bibinfo{year}{2015}), \bibinfo{pages}{191--206}.
\newblock
\showISSN{0163-5964}
\showDOI{%
\url{https://doi.org/10.1145/2786763.2694386}}


\bibitem[\protect\citeauthoryear{David~Kaplan}{David~Kaplan}{2016}]%
        {amd-sme}
\bibfield{author}{\bibinfo{person}{Tom~Woller David~Kaplan, Jeremy~Powell}.}
  \bibinfo{year}{2016}\natexlab{}.
\newblock \bibinfo{booktitle}{{\em White Paper: AMD Memory Encryption}}.
\newblock AMD.
\newblock


\bibitem[\protect\citeauthoryear{Durumeric, Kasten, Adrian, Halderman, Bailey,
  Li, Weaver, Amann, Beekman, Payer, and Paxson}{Durumeric
  et~al\mbox{.}}{2014}]%
        {the-matter-of-heartbleed}
\bibfield{author}{\bibinfo{person}{Zakir Durumeric}, \bibinfo{person}{James
  Kasten}, \bibinfo{person}{David Adrian}, \bibinfo{person}{J.~Alex Halderman},
  \bibinfo{person}{Michael Bailey}, \bibinfo{person}{Frank Li},
  \bibinfo{person}{Nicolas Weaver}, \bibinfo{person}{Johanna Amann},
  \bibinfo{person}{Jethro Beekman}, \bibinfo{person}{Mathias Payer}, {and}
  \bibinfo{person}{Vern Paxson}.} \bibinfo{year}{2014}\natexlab{}.
\newblock \showarticletitle{The Matter of Heartbleed}. In
  \bibinfo{booktitle}{{\em Proceedings of the 2014 Conference on Internet
  Measurement Conference}} {\em (\bibinfo{series}{IMC '14})}.
  \bibinfo{publisher}{ACM}, \bibinfo{address}{New York, NY, USA},
  \bibinfo{pages}{475--488}.
\newblock
\showISBNx{978-1-4503-3213-2}
\showDOI{%
\url{https://doi.org/10.1145/2663716.2663755}}


\bibitem[\protect\citeauthoryear{Erlingsson, Abadi, Vrable, Budiu, and
  Necula}{Erlingsson et~al\mbox{.}}{2006}]%
        {xfi}
\bibfield{author}{\bibinfo{person}{\'{U}lfar Erlingsson},
  \bibinfo{person}{Mart\'{\i}n Abadi}, \bibinfo{person}{Michael Vrable},
  \bibinfo{person}{Mihai Budiu}, {and} \bibinfo{person}{George~C. Necula}.}
  \bibinfo{year}{2006}\natexlab{}.
\newblock \showarticletitle{XFI: Software Guards for System Address Spaces}. In
  \bibinfo{booktitle}{{\em Proceedings of the 7th Symposium on Operating
  Systems Design and Implementation}} {\em (\bibinfo{series}{OSDI '06})}.
  \bibinfo{publisher}{USENIX Association}, \bibinfo{address}{Berkeley, CA,
  USA}, \bibinfo{pages}{75--88}.
\newblock
\showISBNx{1-931971-47-1}
\showURL{%
\url{http://dl.acm.org/citation.cfm?id=1298455.1298463}}


\bibitem[\protect\citeauthoryear{Ford and Cox}{Ford and Cox}{2008}]%
        {vx32}
\bibfield{author}{\bibinfo{person}{Bryan Ford} {and} \bibinfo{person}{Russ
  Cox}.} \bibinfo{year}{2008}\natexlab{}.
\newblock \showarticletitle{Vx32: Lightweight User-level Sandboxing on the
  x86}. In \bibinfo{booktitle}{{\em USENIX 2008 Annual Technical Conference}}
  {\em (\bibinfo{series}{ATC'08})}. \bibinfo{publisher}{USENIX Association},
  \bibinfo{address}{Berkeley, CA, USA}, \bibinfo{pages}{293--306}.
\newblock
\showURL{%
\url{http://dl.acm.org/citation.cfm?id=1404014.1404039}}


\bibitem[\protect\citeauthoryear{Hofmann, Kim, Dunn, Lee, and Witchel}{Hofmann
  et~al\mbox{.}}{2013}]%
        {inktag}
\bibfield{author}{\bibinfo{person}{Owen~S. Hofmann}, \bibinfo{person}{Sangman
  Kim}, \bibinfo{person}{Alan~M. Dunn}, \bibinfo{person}{Michael~Z. Lee}, {and}
  \bibinfo{person}{Emmett Witchel}.} \bibinfo{year}{2013}\natexlab{}.
\newblock \showarticletitle{InkTag: Secure Applications on an Untrusted
  Operating System}.
\newblock \bibinfo{journal}{{\em SIGPLAN Not.\/}} \bibinfo{volume}{48},
  \bibinfo{number}{4} (\bibinfo{date}{March} \bibinfo{year}{2013}),
  \bibinfo{pages}{265--278}.
\newblock
\showISSN{0362-1340}
\showDOI{%
\url{https://doi.org/10.1145/2499368.2451146}}


\bibitem[\protect\citeauthoryear{Hsu, Hoffman, Eugster, and Payer}{Hsu
  et~al\mbox{.}}{2016}]%
        {smv}
\bibfield{author}{\bibinfo{person}{Terry Ching-Hsiang Hsu},
  \bibinfo{person}{Kevin Hoffman}, \bibinfo{person}{Patrick Eugster}, {and}
  \bibinfo{person}{Mathias Payer}.} \bibinfo{year}{2016}\natexlab{}.
\newblock \showarticletitle{Enforcing Least Privilege Memory Views for
  Multithreaded Applications}. In \bibinfo{booktitle}{{\em Proceedings of the
  2016 ACM SIGSAC Conference on Computer and Communications Security}} {\em
  (\bibinfo{series}{CCS '16})}. \bibinfo{publisher}{ACM}, \bibinfo{address}{New
  York, NY, USA}, \bibinfo{pages}{393--405}.
\newblock
\showISBNx{978-1-4503-4139-4}
\showDOI{%
\url{https://doi.org/10.1145/2976749.2978327}}


\bibitem[\protect\citeauthoryear{Inc}{Inc}{2018}]%
        {nginx}
\bibfield{author}{\bibinfo{person}{NGINX Inc}.}
  \bibinfo{year}{2018}\natexlab{}.
\newblock \bibinfo{title}{Nginx}.
\newblock \bibinfo{howpublished}{\url{https://www.nginx.com}}.
  (\bibinfo{year}{2018}).
\newblock
\newblock
\shownote{Last accessed Feb 27 , 2018,.}


\bibitem[\protect\citeauthoryear{{Intel Corporation}}{{Intel
  Corporation}}{2016}]%
        {intel-manual}
\bibfield{author}{\bibinfo{person}{{Intel Corporation}}.}
  \bibinfo{year}{2016}\natexlab{}.
\newblock \bibinfo{booktitle}{{\em {Intel\textsuperscript{\textregistered} 64
  and IA-32 Architectures Software Developer's Manual}}}.
\newblock Number 325462-061US.
\newblock


\bibitem[\protect\citeauthoryear{Isayah R.~(Intel)}{Isayah R.~(Intel)}{2016}]%
        {intel-sgx-arg}
\bibfield{author}{\bibinfo{person}{John M.~(Intel) Isayah R.~(Intel)}.}
  \bibinfo{year}{2016}\natexlab{}.
\newblock \bibinfo{title}{Intel\textsuperscript{\textregistered} SGX Intro:
  Passing Data Between App and Enclave}.
\newblock
  \bibinfo{howpublished}{\url{https://software.intel.com/en-us/articles/sgx-intro-passing-data-between-app-and-enclave}}.
    (\bibinfo{year}{2016}).
\newblock
\newblock
\shownote{Last accessed Feb 27 , 2018,.}


\bibitem[\protect\citeauthoryear{Kamara, Mohassel, and Riva}{Kamara
  et~al\mbox{.}}{2012}]%
        {salus}
\bibfield{author}{\bibinfo{person}{Seny Kamara}, \bibinfo{person}{Payman
  Mohassel}, {and} \bibinfo{person}{Ben Riva}.}
  \bibinfo{year}{2012}\natexlab{}.
\newblock \showarticletitle{Salus: A System for Server-aided Secure Function
  Evaluation}. In \bibinfo{booktitle}{{\em Proceedings of the 2012 ACM
  Conference on Computer and Communications Security}} {\em
  (\bibinfo{series}{CCS '12})}. \bibinfo{publisher}{ACM}, \bibinfo{address}{New
  York, NY, USA}, \bibinfo{pages}{797--808}.
\newblock
\showISBNx{978-1-4503-1651-4}
\showDOI{%
\url{https://doi.org/10.1145/2382196.2382280}}


\bibitem[\protect\citeauthoryear{Kilpatrick}{Kilpatrick}{2003}]%
        {privman}
\bibfield{author}{\bibinfo{person}{Douglas Kilpatrick}.}
  \bibinfo{year}{2003}\natexlab{}.
\newblock \showarticletitle{Privman: A Library for Partitioning Applications.}.
  In \bibinfo{booktitle}{{\em USENIX Annual Technical Conference, FREENIX
  Track}} (2003-09-03). \bibinfo{publisher}{USENIX}, \bibinfo{pages}{273--284}.
\newblock
\showISBNx{1-931971-11-0}
\showURL{%
\url{http://dblp.uni-trier.de/db/conf/usenix/usenix2003f.html#Kilpatrick03}}


\bibitem[\protect\citeauthoryear{Kwon, Dunn, Lee, Hofmann, Xu, and
  Witchel}{Kwon et~al\mbox{.}}{2016}]%
        {sego}
\bibfield{author}{\bibinfo{person}{Youngjin Kwon}, \bibinfo{person}{Alan~M.
  Dunn}, \bibinfo{person}{Michael~Z. Lee}, \bibinfo{person}{Owen~S. Hofmann},
  \bibinfo{person}{Yuanzhong Xu}, {and} \bibinfo{person}{Emmett Witchel}.}
  \bibinfo{year}{2016}\natexlab{}.
\newblock \showarticletitle{Sego: Pervasive Trusted Metadata for Efficiently
  Verified Untrusted System Services}.
\newblock \bibinfo{journal}{{\em SIGOPS Oper. Syst. Rev.\/}}
  \bibinfo{volume}{50}, \bibinfo{number}{2} (\bibinfo{date}{March}
  \bibinfo{year}{2016}), \bibinfo{pages}{277--290}.
\newblock
\showISSN{0163-5980}
\showDOI{%
\url{https://doi.org/10.1145/2954680.2872372}}


\bibitem[\protect\citeauthoryear{Lee, Jang, Jang, Kwak, Choi, Choi, Kim,
  Peinado, and Kang}{Lee et~al\mbox{.}}{2017}]%
        {jaehyuk}
\bibfield{author}{\bibinfo{person}{Jaehyuk Lee}, \bibinfo{person}{Jinsoo Jang},
  \bibinfo{person}{Yeongjin Jang}, \bibinfo{person}{Nohyun Kwak},
  \bibinfo{person}{Yeseul Choi}, \bibinfo{person}{Changho Choi},
  \bibinfo{person}{Taesoo Kim}, \bibinfo{person}{Marcus Peinado}, {and}
  \bibinfo{person}{Brent~ByungHoon Kang}.} \bibinfo{year}{2017}\natexlab{}.
\newblock \showarticletitle{Hacking in Darkness: Return-oriented Programming
  against Secure Enclaves}. In \bibinfo{booktitle}{{\em 26th {USENIX} Security
  Symposium ({USENIX} Security 17)}}. \bibinfo{publisher}{{USENIX}
  Association}, \bibinfo{address}{Vancouver, BC}, \bibinfo{pages}{523--539}.
\newblock
\showISBNx{978-1-931971-40-9}
\showURL{%
\url{https://www.usenix.org/conference/usenixsecurity17/technical-sessions/presentation/lee-jaehyuk}}


\bibitem[\protect\citeauthoryear{Li, McCune, Newsome, Perrig, Baker, and
  Drewry}{Li et~al\mbox{.}}{2014}]%
        {minibox}
\bibfield{author}{\bibinfo{person}{Yanlin Li}, \bibinfo{person}{Jonathan
  McCune}, \bibinfo{person}{James Newsome}, \bibinfo{person}{Adrian Perrig},
  \bibinfo{person}{Brandon Baker}, {and} \bibinfo{person}{Will Drewry}.}
  \bibinfo{year}{2014}\natexlab{}.
\newblock \showarticletitle{MiniBox: A Two-Way Sandbox for x86 Native Code}. In
  \bibinfo{booktitle}{{\em 2014 USENIX Annual Technical Conference (USENIX ATC
  14)}}. \bibinfo{publisher}{USENIX Association},
  \bibinfo{address}{Philadelphia, PA}, \bibinfo{pages}{409--420}.
\newblock
\showISBNx{978-1-931971-10-2}
\showURL{%
\url{https://www.usenix.org/conference/atc14/technical-sessions/presentation/li_yanlin}}


\bibitem[\protect\citeauthoryear{Limited}{Limited}{2009}]%
        {arm-trustzone}
\bibfield{author}{\bibinfo{person}{ARM Limited}.}
  \bibinfo{year}{2009}\natexlab{}.
\newblock \bibinfo{title}{Building a Secure System using TrustZone®
  Technolog}.
\newblock
  \bibinfo{howpublished}{\url{http://infocenter.arm.com/help/topic/com.arm.doc.prd29-genc-009492c/PRD29-GENC-009492C_trustzone_security_whitepaper.pdf}}.
    (\bibinfo{year}{2009}).
\newblock


\bibitem[\protect\citeauthoryear{Liu, Zhou, Chen, Chen, and Xia}{Liu
  et~al\mbox{.}}{2015}]%
        {secage}
\bibfield{author}{\bibinfo{person}{Yutao Liu}, \bibinfo{person}{Tianyu Zhou},
  \bibinfo{person}{Kexin Chen}, \bibinfo{person}{Haibo Chen}, {and}
  \bibinfo{person}{Yubin Xia}.} \bibinfo{year}{2015}\natexlab{}.
\newblock \showarticletitle{Thwarting Memory Disclosure with Efficient
  Hypervisor-enforced Intra-domain Isolation}. In \bibinfo{booktitle}{{\em
  Proceedings of the 22Nd ACM SIGSAC Conference on Computer and Communications
  Security}} {\em (\bibinfo{series}{CCS '15})}. \bibinfo{publisher}{ACM},
  \bibinfo{address}{New York, NY, USA}, \bibinfo{pages}{1607--1619}.
\newblock
\showISBNx{978-1-4503-3832-5}
\showDOI{%
\url{https://doi.org/10.1145/2810103.2813690}}


\bibitem[\protect\citeauthoryear{McCamant and Morrisett}{McCamant and
  Morrisett}{2006}]%
        {sfi-cisc}
\bibfield{author}{\bibinfo{person}{Stephen McCamant} {and}
  \bibinfo{person}{Greg Morrisett}.} \bibinfo{year}{2006}\natexlab{}.
\newblock \showarticletitle{Evaluating SFI for a CISC Architecture}. In
  \bibinfo{booktitle}{{\em Proceedings of the 15th Conference on USENIX
  Security Symposium - Volume 15}} {\em (\bibinfo{series}{USENIX-SS'06})}.
  \bibinfo{publisher}{USENIX Association}, \bibinfo{address}{Berkeley, CA,
  USA}, Article \bibinfo{articleno}{15}.
\newblock
\showURL{%
\url{http://dl.acm.org/citation.cfm?id=1267336.1267351}}


\bibitem[\protect\citeauthoryear{McCune, Li, Qu, Zhou, Datta, Gligor, and
  Perrig}{McCune et~al\mbox{.}}{2010}]%
        {trustvisor}
\bibfield{author}{\bibinfo{person}{J.~M. McCune}, \bibinfo{person}{Y. Li},
  \bibinfo{person}{N. Qu}, \bibinfo{person}{Z. Zhou}, \bibinfo{person}{A.
  Datta}, \bibinfo{person}{V. Gligor}, {and} \bibinfo{person}{A. Perrig}.}
  \bibinfo{year}{2010}\natexlab{}.
\newblock \showarticletitle{TrustVisor: Efficient TCB Reduction and
  Attestation}. In \bibinfo{booktitle}{{\em 2010 IEEE Symposium on Security and
  Privacy}}. \bibinfo{pages}{143--158}.
\newblock
\showISSN{1081-6011}
\showDOI{%
\url{https://doi.org/10.1109/SP.2010.17}}


\bibitem[\protect\citeauthoryear{Morrisett, Tan, Tassarotti, Tristan, and
  Gan}{Morrisett et~al\mbox{.}}{2012}]%
        {rocksalt}
\bibfield{author}{\bibinfo{person}{Greg Morrisett}, \bibinfo{person}{Gang Tan},
  \bibinfo{person}{Joseph Tassarotti}, \bibinfo{person}{Jean-Baptiste Tristan},
  {and} \bibinfo{person}{Edward Gan}.} \bibinfo{year}{2012}\natexlab{}.
\newblock \showarticletitle{RockSalt: Better, Faster, Stronger SFI for the
  x86}. In \bibinfo{booktitle}{{\em Proceedings of the 33rd ACM SIGPLAN
  Conference on Programming Language Design and Implementation}} {\em
  (\bibinfo{series}{PLDI '12})}. \bibinfo{publisher}{ACM},
  \bibinfo{address}{New York, NY, USA}, \bibinfo{pages}{395--404}.
\newblock
\showISBNx{978-1-4503-1205-9}
\showDOI{%
\url{https://doi.org/10.1145/2254064.2254111}}


\bibitem[\protect\citeauthoryear{OpenBSD}{OpenBSD}{2017}]%
        {libressl}
\bibfield{author}{\bibinfo{person}{OpenBSD}.} \bibinfo{year}{2017}\natexlab{}.
\newblock \bibinfo{title}{LibreSSL}.
\newblock \bibinfo{howpublished}{\url{http://www.libressl.org}}.
  (\bibinfo{year}{2017}).
\newblock
\newblock
\shownote{Last accessed Feb 27 , 2018,.}


\bibitem[\protect\citeauthoryear{Organization}{Organization}{2018}]%
        {kernel-web}
\bibfield{author}{\bibinfo{person}{Linux~Kernel Organization}.}
  \bibinfo{year}{2018}\natexlab{}.
\newblock \bibinfo{title}{The Linux Kernel Archives}.
\newblock \bibinfo{howpublished}{\url{https://www.kernel.org}}.
  (\bibinfo{year}{2018}).
\newblock
\newblock
\shownote{Last accessed April 2 , 2018,.}


\bibitem[\protect\citeauthoryear{Provos, Friedl, and Honeyman}{Provos
  et~al\mbox{.}}{2003}]%
        {ppe}
\bibfield{author}{\bibinfo{person}{Niels Provos}, \bibinfo{person}{Markus
  Friedl}, {and} \bibinfo{person}{Peter Honeyman}.}
  \bibinfo{year}{2003}\natexlab{}.
\newblock \showarticletitle{Preventing Privilege Escalation}. In
  \bibinfo{booktitle}{{\em Proceedings of the 12th Conference on USENIX
  Security Symposium - Volume 12}} {\em (\bibinfo{series}{SSYM'03})}.
  \bibinfo{publisher}{USENIX Association}, \bibinfo{address}{Berkeley, CA,
  USA}, \bibinfo{pages}{16--16}.
\newblock
\showURL{%
\url{http://dl.acm.org/citation.cfm?id=1251353.1251369}}


\bibitem[\protect\citeauthoryear{Sehr, Muth, Biffle, Khimenko, Pasko, Schimpf,
  Yee, and Chen}{Sehr et~al\mbox{.}}{2010}]%
        {nacl-ext}
\bibfield{author}{\bibinfo{person}{David Sehr}, \bibinfo{person}{Robert Muth},
  \bibinfo{person}{Cliff Biffle}, \bibinfo{person}{Victor Khimenko},
  \bibinfo{person}{Egor Pasko}, \bibinfo{person}{Karl Schimpf},
  \bibinfo{person}{Bennet Yee}, {and} \bibinfo{person}{Brad Chen}.}
  \bibinfo{year}{2010}\natexlab{}.
\newblock \showarticletitle{Adapting Software Fault Isolation to Contemporary
  CPU Architectures}. In \bibinfo{booktitle}{{\em Proceedings of the 19th
  USENIX Conference on Security}} {\em (\bibinfo{series}{USENIX Security'10})}.
  \bibinfo{publisher}{USENIX Association}, \bibinfo{address}{Berkeley, CA,
  USA}, \bibinfo{pages}{1--1}.
\newblock
\showISBNx{888-7-6666-5555-4}
\showURL{%
\url{http://dl.acm.org/citation.cfm?id=1929820.1929822}}


\bibitem[\protect\citeauthoryear{Wahbe, Lucco, Anderson, and Graham}{Wahbe
  et~al\mbox{.}}{1993}]%
        {sfi}
\bibfield{author}{\bibinfo{person}{Robert Wahbe}, \bibinfo{person}{Steven
  Lucco}, \bibinfo{person}{Thomas~E. Anderson}, {and} \bibinfo{person}{Susan~L.
  Graham}.} \bibinfo{year}{1993}\natexlab{}.
\newblock \showarticletitle{Efficient Software-based Fault Isolation}. In
  \bibinfo{booktitle}{{\em Proceedings of the Fourteenth ACM Symposium on
  Operating Systems Principles}} {\em (\bibinfo{series}{SOSP '93})}.
  \bibinfo{publisher}{ACM}, \bibinfo{address}{New York, NY, USA},
  \bibinfo{pages}{203--216}.
\newblock
\showISBNx{0-89791-632-8}
\showDOI{%
\url{https://doi.org/10.1145/168619.168635}}


\bibitem[\protect\citeauthoryear{Wang, Xiong, and Liu}{Wang
  et~al\mbox{.}}{2015}]%
        {arbiter}
\bibfield{author}{\bibinfo{person}{Jun Wang}, \bibinfo{person}{Xi Xiong}, {and}
  \bibinfo{person}{Peng Liu}.} \bibinfo{year}{2015}\natexlab{}.
\newblock \showarticletitle{Between Mutual Trust and Mutual Distrust: Practical
  Fine-grained Privilege Separation in Multithreaded Applications}. In
  \bibinfo{booktitle}{{\em Proceedings of the 2015 USENIX Conference on Usenix
  Annual Technical Conference}} {\em (\bibinfo{series}{USENIX ATC '15})}.
  \bibinfo{publisher}{USENIX Association}, \bibinfo{address}{Berkeley, CA,
  USA}, \bibinfo{pages}{361--373}.
\newblock
\showISBNx{978-1-931971-225}
\showURL{%
\url{http://dl.acm.org/citation.cfm?id=2813767.2813794}}


\bibitem[\protect\citeauthoryear{Wheeler}{Wheeler}{2014}]%
        {preventing-heartbleed}
\bibfield{author}{\bibinfo{person}{D.~A. Wheeler}.}
  \bibinfo{year}{2014}\natexlab{}.
\newblock \showarticletitle{Preventing Heartbleed}.
\newblock \bibinfo{journal}{{\em Computer\/}} \bibinfo{volume}{47},
  \bibinfo{number}{8} (\bibinfo{date}{Aug} \bibinfo{year}{2014}),
  \bibinfo{pages}{80--83}.
\newblock
\showISSN{0018-9162}
\showDOI{%
\url{https://doi.org/10.1109/MC.2014.217}}


\bibitem[\protect\citeauthoryear{Yang and Shin}{Yang and Shin}{2008}]%
        {yang}
\bibfield{author}{\bibinfo{person}{Jisoo Yang} {and} \bibinfo{person}{Kang~G.
  Shin}.} \bibinfo{year}{2008}\natexlab{}.
\newblock \showarticletitle{Using Hypervisor to Provide Data Secrecy for User
  Applications on a Per-page Basis}. In \bibinfo{booktitle}{{\em Proceedings of
  the Fourth ACM SIGPLAN/SIGOPS International Conference on Virtual Execution
  Environments}} {\em (\bibinfo{series}{VEE '08})}. \bibinfo{publisher}{ACM},
  \bibinfo{address}{New York, NY, USA}, \bibinfo{pages}{71--80}.
\newblock
\showISBNx{978-1-59593-796-4}
\showDOI{%
\url{https://doi.org/10.1145/1346256.1346267}}


\bibitem[\protect\citeauthoryear{Yee, Sehr, Dardyk, Chen, Muth, Ormandy,
  Okasaka, Narula, and Fullagar}{Yee et~al\mbox{.}}{2009}]%
        {nacl}
\bibfield{author}{\bibinfo{person}{Bennet Yee}, \bibinfo{person}{David Sehr},
  \bibinfo{person}{Gregory Dardyk}, \bibinfo{person}{J.~Bradley Chen},
  \bibinfo{person}{Robert Muth}, \bibinfo{person}{Tavis Ormandy},
  \bibinfo{person}{Shiki Okasaka}, \bibinfo{person}{Neha Narula}, {and}
  \bibinfo{person}{Nicholas Fullagar}.} \bibinfo{year}{2009}\natexlab{}.
\newblock \showarticletitle{Native Client: A Sandbox for Portable, Untrusted
  x86 Native Code}. In \bibinfo{booktitle}{{\em Proceedings of the 2009 30th
  IEEE Symposium on Security and Privacy}} {\em (\bibinfo{series}{SP '09})}.
  \bibinfo{publisher}{IEEE Computer Society}, \bibinfo{address}{Washington, DC,
  USA}, \bibinfo{pages}{79--93}.
\newblock
\showISBNx{978-0-7695-3633-0}
\showDOI{%
\url{https://doi.org/10.1109/SP.2009.25}}


\end{thebibliography}

\end{document}